\documentclass[aps,prb,superscriptaddress,twocolumn,notitlepage,10pt,longbibliography]{revtex4-2}
\usepackage[english]{babel}

\usepackage{amssymb,amsmath,amsthm,amsfonts,mathtools,mathrsfs,chemformula,braket}
\usepackage{graphicx,subcaption,float,comment,bm,enumitem,cancel,xr-hyper,booktabs,threeparttable,multirow}
\usepackage{soul,xcolor,colortbl}
\usepackage{ragged2e} 

\makeatletter
\long\def\@makecaption#1#2{%
  \vskip\abovecaptionskip
  \small
  \setbox\@tempboxa\hbox{#1\quad #2}%
  \ifdim\wd\@tempboxa>\linewidth
    \parbox{\linewidth}{\justifying #1\quad #2}\par
  \else
    \hbox to\linewidth{\hfil\box\@tempboxa\hfil}%
  \fi
  \vskip\belowcaptionskip
}
\makeatother

\usepackage[colorlinks]{hyperref}
\usepackage{cleveref}

\newcommand{\eq}[1]{Eq.~\hyperref[eq:#1]{(\ref*{eq:#1})}}
\renewcommand{\sec}[1]{\hyperref[sec:#1]{Section~\ref*{sec:#1}}}
\newcommand{\app}[1]{\hyperref[app:#1]{Appendix~\ref*{app:#1}}}
\newcommand{\tab}[1]{\hyperref[tab:#1]{Table~\ref*{tab:#1}}}
\newcommand{\fig}[1]{\hyperref[fig:#1]{Figure~\ref*{fig:#1}}}
\newcommand{\figa}[2]{\hyperref[fig:#1]{Figure~\ref*{fig:#1}#2}}
\newcommand{\figx}[2]{\hyperref[fig:#1]{Figure~\ref*{fig:#1}(#2)}}
\newcommand{\thm}[1]{\hyperref[thm:#1]{Theorem~\ref*{thm:#1}}}
\newcommand{\lem}[1]{\hyperref[lem:#1]{Lemma~\ref*{lem:#1}}}
\newcommand{\cor}[1]{\hyperref[cor:#1]{Corollary~\ref*{cor:#1}}}
\newcommand{\defn}[1]{\hyperref[def:#1]{Definition~\ref*{def:#1}}}
\newcommand{\alg}[1]{\hyperref[alg:#1]{Algorithm~\ref*{alg:#1}}}

\newcommand{\be}{\begin{equation}}
\newcommand{\ee}{\end{equation}}
\newcommand{\ba}{\begin{eqnarray}}
\newcommand{\ea}{\end{eqnarray}}

\def\ket#1{\mathinner{|{#1}\rangle}}

\makeatletter
\renewcommand*\env@matrix[1][\arraystretch]{%
	\edef\arraystretch{#1}%
	\hskip -\arraycolsep
	\let\@ifnextchar\new@ifnextchar
	\array{*\c@MaxMatrixCols c}}
\makeatother
\makeatletter
\newcommand*{\addFileDependency}[1]{
\typeout{(#1)}
\@addtofilelist{#1}
\IfFileExists{#1}{}{\typeout{No file #1.}}
}\makeatother

\newcommand{\Columbia}{\affiliation{Department of Chemistry, Columbia University, New York, New York 10027, USA}}

\begin{document}
\title{Study of the triangular-lattice Hubbard model with constrained-path quantum Monte Carlo}
\date{\today}

\author{Shu Fay Ung}
\email{su2254@columbia.edu}
\thanks{These authors contributed equally to this work.}
\Columbia

\author{Ankit Mahajan}
\email{ankitmahajan76@gmail.com}
\thanks{These authors contributed equally to this work.}
\Columbia

\author{David R. Reichman}
\email{drr2103@columbia.edu}
\Columbia

\begin{abstract}
    We benchmark constrained-path Monte Carlo (CPMC) on the triangular-lattice Hubbard model for several fillings and $U$ values and show that symmetry-adapted trial wave functions substantially improve quantitative accuracy. Away from half-filling, simple free-electron-based trials that preserve the ground state symmetry yield energy deviations $\lesssim 1\%$ from exact diagonalization and density matrix renormalization group results. At half-filling, strong frustration in the intermediate to large $U$ regimes necessitates symmetry-projected trials to reach comparable accuracy, where both free-electron and symmetry-broken Hartree-Fock trials incur substantial constraint bias. Since the computational cost of CPMC with symmetry projection scales polynomially with system size, our results motivate its use as a practical route for studying competing ground states in strongly correlated, frustrated systems.
\end{abstract}

\maketitle

\section{Introduction}
The Hubbard model has long served as a central paradigm for understanding the emergence of strongly correlated phenomena, ranging from magnetic order and metal-insulator transitions to superconductivity \cite{qin_hubbard_2022}. In recent years, the discovery of triangular moiré superlattices in graphene and transition-metal dichalcogenide (TMD) heterostructures has reinvigorated interest in the triangular-lattice Hubbard model \cite{pan_quantum_2020,zang_hartree-fock_2021,morales-duran_nonlocal_2022,zhou_quantum_2024,kumar_origin_2025,tang_simulation_2025}, which has been shown to capture the low-energy physics of TMD heterobilayers \cite{wu_hubbard_2018}. Extensive study of the model was previously spurred by the discovery of superconductivity and spin-liquid states in organic salts \cite{shimizu_spin_2003,itou_quantum_2008,yamashita_thermodynamic_2008,yamashita_thermal-transport_2009,itou_instability_2010,yamashita_highly_2010,yamashita_gapless_2011}, as it offers a minimal framework for interpreting experiments on strongly correlated, frustrated systems. 

Theoretically, the triangular-lattice Hubbard model is an important benchmark for many-body computational methods due to the complex interplay of geometric frustration and electronic correlations. At half-filling in the limit of large on-site repulsion $U$ (\textit{i.e.} $U \to \infty$), the model maps onto the Heisenberg model, which was originally proposed by Anderson to host a quantum spin-liquid ground state \cite{anderson_resonating_1973}. Subsequent numerical studies find that the ground state retains the $120^\circ$ Néel order, albeit with a significantly reduced order parameter compared to the classical prediction \cite{huse_simple_1988,white_neel_2007}. Introducing additional frustration, for example via next-nearest-neighbor interactions or ring-exchange terms, has been shown to melt the magnetic order and stabilize spin-liquid phases \cite{motrunich_variational_2005,sheng_spin_2009,kaneko_gapless_2014,zhu_spin_2015,cookmeyer_four-spin_2021}. 

In the limit of small $U$, the ground state is metallic; the intermediate-$U$ regime, where charge fluctuations remain, is still under intense debate. The existence of a non-magnetic insulating phase has been predicted by several computational methods including the path-integral renormalization group \cite{morita_nonmagnetic_2002,yoshioka_quantum_2009}, variational cluster approximation \cite{sahebsara_hubbard_2008}, exact diagonalization (ED) \cite{koretsune_exact_2007,kokalj_thermodynamics_2013}, and the ladder dual-fermion approach \cite{li_competing_2014}. Recent DMRG studies \cite{szasz_chiral_2020,szasz_phase_2021,chen_quantum_2022} provide evidence that the intermediate phase is a chiral spin liquid, while earlier DMRG results \cite{shirakawa_ground-state_2017} challenged this and variational Monte Carlo studies \cite{tocchio_magnetic_2020,tocchio_hubbard_2021} instead suggest the absence of a spin-liquid phase. A recent ``multi-method, multi-messenger" approach \cite{wietek_mott_2021} also finds competing chiral and magnetic orders at intermediate $U$.

In the doped regime relevant to many triangular-lattice materials, DMRG calculations have revealed a rich phase diagram comprising chiral metallic, spin density wave, and enhanced pair-pair correlation phases as doping and interaction strength are varied \cite{zhu_doped_2022}. Significant effort has focused on superconductivity and its pairing symmetries \cite{chen_unconventional_2013,gomes_coulomb-enhanced_2016,de_silva_coulomb_2016,venderley_density_2019}, driven by the discovery of superconductivity in organic salts \cite{yamashita_thermodynamic_2008,yamashita_thermal-transport_2009} and cobaltates \cite{takada_superconductivity_2003}. 

Given the complexity of the phase diagram, it is valuable to compare results across complementary numerical approaches. Quantum Monte Carlo (QMC) techniques have played a central role in elucidating the Hubbard model, especially on the square lattice, where the fermion sign problem can be eliminated at half-filling \cite{loh_sign_1990}. Due to its non-bipartite nature, one generically encounters a sign problem on the triangular lattice, including at half-filling. Constrained-path Monte Carlo (CPMC), a projector QMC method, is particularly well-suited for the Hubbard interaction and has been extensively benchmarked on the square lattice \cite{zhang_constrained_1995,zhang_constrained_1999,chang_spatially_2008,shi_symmetry_2013,shi_symmetry-projected_2014,simons_collaboration_on_the_many-electron_problem_solutions_2015,qin_coupling_2016,qin_benchmark_2016,zheng_stripe_2017}. In contrast, its performance has not been systematically assessed to a comparable extent on the triangular lattice, with only a few recent studies \cite{yang_metal-insulator_2023,zampronio_chiral_2023}.

In this paper, we present a detailed CPMC study of the triangular-lattice Hubbard model at several fillings. We show that the use of appropriate symmetry-adapted trial wave functions is important for obtaining accurate results, especially at half-filling where strong frustration makes the problem particularly challenging. Our results help clarify the ground-state properties of the triangular-lattice Hubbard model and demonstrate the effectiveness of symmetry projection in CPMC simulations of frustrated systems. This provides a potentially powerful route to exploring emergent states of the model in a controlled manner as the thermodynamic limit is approached.

The paper is organized as follows. We first introduce the model, the CPMC method, and our symmetry-adapted trial states in Sec. \ref{methods}. We then present results in Sec. \ref{results} and conclude with a discussion in Sec. \ref{discussion}.

\section{Model and methods} \label{methods}
\subsection{The Hubbard model}
\begin{figure*}[t]
  \centering
  \begin{subfigure}[t]{0.54\textwidth}
    \centering
    \includegraphics[width=\linewidth]{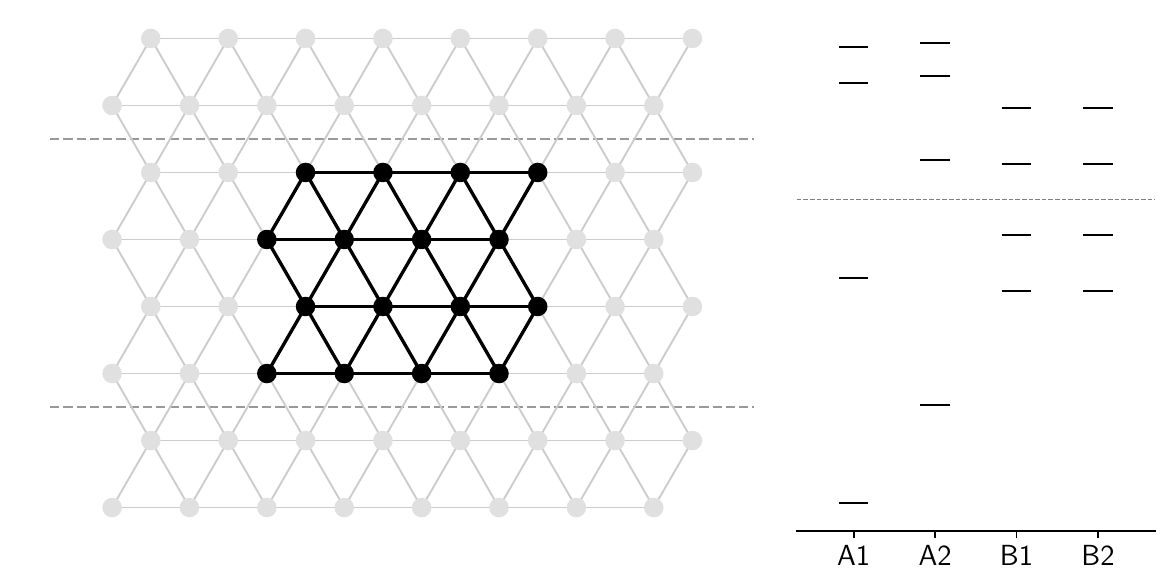}
    \caption{XC $4\times4$}
  \end{subfigure}\hfill
  \begin{subfigure}[t]{0.45\textwidth}
    \centering
    \includegraphics[width=\linewidth]{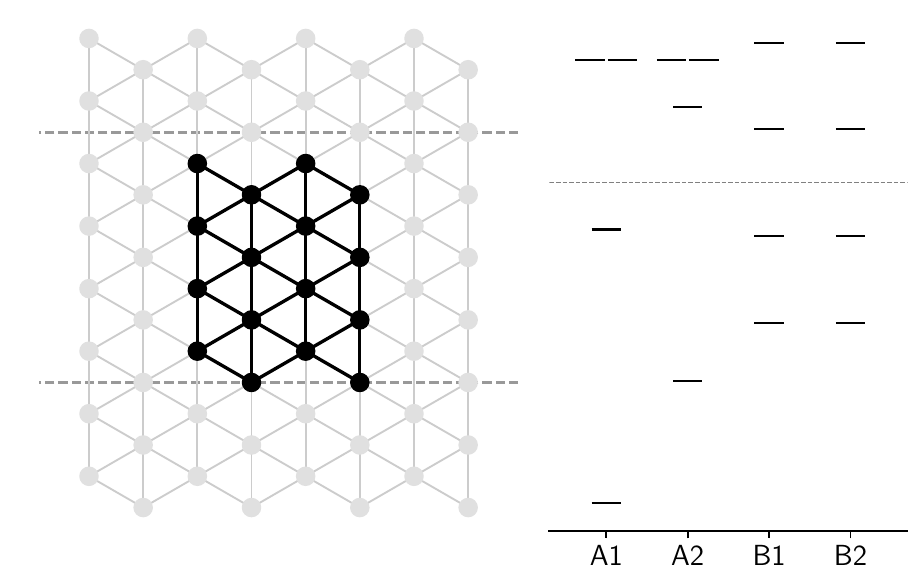}
    \caption{YC $4\times4$}
  \end{subfigure}
  \caption{Triangular-lattice geometries on XC (a) and YC (b) cylinders. In each subfigure, an illustration of the lattice is shown on the left, while the corresponding one-electron spectra (eigenvalues of $\hat{H}_1$) grouped by $D_2$ irreps is shown on the right. The black outline shows a $4 \times 4$ region with periodic boundaries across the dashed lines. Note that while the symmetry groups of the XC $4 \times 4$ and YC $4 \times 4$ lattices are $D_4$ and $D_8$, respectively, one can label the orbitals using irreps of the $D_2$ subgroup (see Appendix~\ref{app:xc4_symm} for details).}
  \label{fig:lattices}
\end{figure*}

The Hubbard model is described by the lattice Hamiltonian
\begin{align}
    \hat{H} = \underbrace{-t \sum_\sigma \sum_{\braket{ij}} \left[ \hat{c}^\dagger_{i\sigma} \hat{c}_{j\sigma} + \text{h.c.} \right]}_{\hat{H}_1} + \underbrace{U \sum_{i} \hat{n}_{i \uparrow} \hat{n}_{i \downarrow}}_{\hat{H}_2},
\end{align}
where $\hat{c}_{i, \sigma}^\dagger \left( \hat{c}_{i, \sigma}\right)$ is the creation (annihilation) operator for an electron at site $i$ and spin $\sigma \in \{\uparrow, \downarrow\}$, $\hat{n}_{i\sigma} = \hat{c}_{i\sigma}^\dagger \hat{c}_{i\sigma}$ is the number operator, $t \geq 0$ is the hopping parameter, $U \geq 0$ is the interaction strength, and $\sum_{\braket{ij}}$ enumerates nearest-neighbor sites only. We define the filling factor $\nu = (N_{\uparrow} + N_{\downarrow}) / N_s$, where $N_\sigma$ denotes the number of spin-$\sigma$ electrons and $N_s$ the number of sites. In this work, we focus on ground states in the $\hat{S}_z = 0$ spin sector (\textit{i.e.} $N_{\uparrow} = N_{\downarrow}$) and the hole-doped to half-filling regime $\nu \leq 1$. We set $t = 1$ in all calculations reported.

We perform our CPMC calculations on finite cylinders with periodic (open) boundary conditions and $N_y$ ($N_x$) sites along the $y$- ($x$-) direction, giving a total number of $N_s = N_x N_y$ sites. This geometry facilitates direct comparison with DMRG results and reduces open-shell effects associated with degeneracies in the one-electron spectrum, although it is not strictly required for CPMC. For the triangular lattice, two commonly considered configurations are the XC$n$ and YC$n$ cylinders, where $n$ denotes the number of bonds along the $y$-direction \cite{shirakawa_ground-state_2017}. Illustrative examples are depicted in Fig. \ref{fig:lattices}.

\subsection{Constrained-path Monte Carlo}
CPMC is a projector quantum Monte Carlo method that filters out the ground state $\ket{\Psi_0}$ from an initial state $\ket{\Phi^{(0)}}$ (such that $\braket{\Psi_0 | \Phi^{(0)}} \neq 0$) via imaginary-time evolution,
\begin{align}
    \ket{\Psi_0} \propto \lim_{\tau \to \infty} e^{-\tau \hat{H}} \ket{\Phi^{(0)}}.
\end{align}
In practice, the long-time propagator is first discretized into $n$ time steps $\Delta \tau = \tau / n$ and Trotterized to obtain the short-time propagator
\begin{align}
    e^{-\Delta \tau \hat{H}} &= e^{-\Delta \tau \hat{H}_1/2} e^{-\Delta \tau \hat{H}_2} e^{-\Delta \tau \hat{H}_1/2} + \mathcal{O}(\Delta \tau^3). \label{prop}
\end{align}
The two-body propagator $e^{-\Delta \tau \hat{H}_2}$ is then expressed via a Hubbard-Stratonovich (HS) transformation as a linear combination of one-body propagators coupled to fluctuating auxiliary fields. For the Hubbard interaction, this is achieved with Hirsch's discrete formulation (we use the spin decomposition) \cite{hirsch_discrete_1983,hirsch_two-dimensional_1985}:
\begin{align}
    e^{-\Delta \tau U \hat{n}_{i\uparrow} \hat{n}_{i\downarrow}} = e^{-\Delta \tau U (\hat{n}_{i\uparrow} + \hat{n}_{i\downarrow}) / 2} \sum_{x_i = \pm 1} p(x_i) e^{\gamma x_i (\hat{n}_{i\uparrow} - \hat{n}_{i\downarrow})}, \label{hs}
\end{align}
where $\gamma$ is determined by $\cosh{\gamma} = e^{\Delta \tau U / 2}$ and $p(x_i)$ is a discrete probability density function over $x_i$. The projection of the ground state is hence implemented using Monte Carlo sampling of the auxiliary fields $\{x_i\}$ as an open-ended random walk starting from $\ket{\Phi^{(0)}}$ \cite{zhang_constrained_1995,zhang_constrained_1999}.

The state $\ket{\Phi^{(k)}}$ at time $k \Delta \tau$ is represented as an ensemble of Slater determinants $\ket{\phi^{(k)}_i}$ (\textit{i.e.} walkers) with weights $w_i$,
\begin{align}
    \ket{\Phi^{(k)}} = \sum_i w_i \ket{\phi^{(k)}_i}. \label{state}
\end{align}
Since the propagator \eqref{prop} is expressed only in terms of one-body operators in the exponent, Thouless' theorem \cite{thouless_stability_1960,thouless_vibrational_1961} implies that the evolution of walkers is constrained to the manifold of Slater determinants. Due to the symmetry between the ground state $\ket{\Psi_0}$ and its negative $-\ket{\Psi_0}$, the approximate propagation by \eqref{hs} can cause walkers to cross the nodal surface defined by $\braket{\Psi_0 | \phi_i} = 0$ and acquire a minus sign, resulting in the sign problem \cite{loh_sign_1990,troyer_computational_2005}. CPMC eliminates this issue by imposing a constraint on the sign of walkers via a trial wave function $\ket{\psi_T}$, reducing the sample complexity from exponential to polynomial scaling in the system size \cite{lee_twenty_2022}. However, this introduces a bias that depends on the quality of $\ket{\psi_T}$ in approximating $\ket{\Psi_0}$. Choices of $\ket{\psi_T}$ range from the simplistic free-electron (FE) and Hartree-Fock (HF) Slater determinants to more sophisticated trials that incorporate a self-consistent constraint \cite{qin_coupling_2016,zheng_stripe_2017,xu_stripes_2022,qin_self-consistent_2023} or use symmetry-projection \cite{rodriguez-guzman_symmetry-projected_2012,shi_symmetry-projected_2014}. In practice, the application of \eqref{hs} also leverages importance sampling to improve efficiency and reduce fluctuations in $w_i$.

Given a sufficiently equilibrated $\ket{\Phi^{(k)}}$, the expectation value of observables $\hat{O}$ can be estimated using the mixed estimator
\begin{align}
    \braket{\hat{O}} = \frac{\sum_i w_i \ O_L[\phi_i^{(k)}]}{\sum_i w_i}, \quad O_L[\phi_i^{(k)}] = \frac{\braket{\psi_T | \hat{O} | \phi^{(k)}_i}}{\braket{\psi_T | \phi^{(k)}_i}}, \label{estimator}
\end{align}
where $O_L[\phi_i^{(k)}]$ is the local estimator. For $\hat{O}$ that commute with $\hat{H}$ (such as $\hat{O} = \hat{H}$ so $\braket{\hat{H}}$ is the energy), the mixed estimator is exact if $\ket{\psi_T} = \ket{\Psi_0}$; it is biased otherwise. In the latter case, back-propagation \cite{motta_computation_nodate,purwanto_quantum_2004,zhang_constrained_1997} or extrapolations \cite{purwanto_quantum_2004,mahajan_selected_2022} may be applied to obtain unbiased estimates.

\subsection{Symmetry-projected trial wave functions}\label{sec:symm_proj_trial}
Previous work on the square lattice Hubbard model \cite{shi_symmetry_2013,shi_symmetry-projected_2014} has shown that symmetry properties in the trial wave function play a significant role in reducing the constraint bias. In particular, \emph{deliberately} breaking and restoring symmetries in the wave function \cite{ring_nuclear_1980,fukutome_unrestricted_1981,Scuseria2011} yields a hierarchy of trial wave functions with increasing quality. This includes continuous symmetries such as SU(2) spin-rotation, as well as discrete symmetries such as lattice space-group operations and complex-conjugation.

Starting from an arbitrary broken-symmetry HF determinant, one may restore these symmetries by applying the proper projection operators. 
Spin symmetry implies that the wave function is independent of the choice of spin-quantization axis, leading to the projector \cite{percus_exact_1962,lefebvre_etudes_1969}
\begin{align}
    \hat{\mathcal{P}}^s_{mm'} &= \frac{2s+1}{8\pi^2} \int d\Omega \ D^s_{mm'}(\Omega)^* \hat{R}(\Omega),
\end{align}
where $D^{s}_{mm'}(\Omega) = \braket{s; m | \hat{R}(\Omega) | s; m'}$ is the Wigner $D$-matrix and $\hat{R}(\Omega) = e^{-i\alpha \hat{S}_z} e^{-i\beta \hat{S}_y} e^{-i\gamma \hat{S}_z}$ is the rotation operator parametrized by Euler angles $\Omega = (\alpha, \beta, \gamma)$ in spin space. By restoring spin-rotational invariance, the wave function is forced to be an eigenstate of $\hat{S}^2$ \cite{Scuseria2011}.

For a Hamiltonian invariant under a spatial symmetry group $\mathcal{G}$, its eigenstates $\ket{\Psi}$ can be chosen to transform according to irreducible representations (irreps) of $\mathcal{G}$. To recover spatial symmetries, we hence use the projector onto irrep $\alpha$ of $\mathcal{G}$ \cite{shi_symmetry-projected_2014} given by
\begin{align}
    \hat{\mathcal{P}}_{\alpha} = \frac{d_{\alpha}}{|\mathcal{G}|} \sum_{g \in \mathcal{G}} \chi_{\alpha}(g)^* \hat{R}(g),
\end{align}
where $\hat{R}(g)$ is the symmetry operator labeled by the group element $g$, $\chi_{\alpha}$ is the character of $g$ in the irrep, $d_\alpha$ is the irrep dimension, and $|\mathcal{G}|$ is the order of the group.

Finally, in cases where the Hamiltonian is real in some basis, $\hat{H}$ satisfies \([\hat{H}, \hat{K}] = 0\) where $\hat{K}$ is the complex-conjugation operator. For a non-degenerate eigenstate $\ket{\Psi}$, this implies $\hat{K} \ket{\Psi} = e^{i\phi} \ket{\Psi}$ for some real $\phi$. We can always choose a gauge in which $\phi = 0$. For a general state that does not satisfy this condition, we can enforce it by defining the projector
\begin{align}
    \hat{\mathcal{P}}_K = \frac{1}{2}(\hat{I} + \hat{K}),
\end{align}
since $\hat{K}^2 = \hat{I}$ and 
\begin{align}
    \hat{K} \hat{\mathcal{P}}_K \ket{\Psi} = \frac{1}{2}(\hat{K} + \hat{I}) \ket{\Psi} = \hat{\mathcal{P}}_K \ket{\Psi}.
\end{align}
We note that while breaking complex-conjugation symmetry requires using complex-valued states $\ket{\Psi}$ (in the site basis), the symmetry-restored state $\hat{\mathcal{P}}_K \ket{\Psi}$ is real-valued. Consequently, the overlap ratio computed with $\hat{\mathcal{P}}_K \ket{\Psi}$ is also real-valued and we only encounter the sign and not phase problem.

\begin{table*}[t]
    \setlength{\belowcaptionskip}{10pt}
    \caption{
        Ground state energies per site obtained from CPMC and ED for the XC $4 \times 4$ lattice at fillings $\nu \leq 1$. FE trials can yield accuracies comparable to symmetry-projected trial states, especially away from half-filling. The relative (rel.) error is computed as $(\varepsilon_\text{CPMC} - \varepsilon_\text{exact}) / |\varepsilon_\text{exact}| \times 100$. Tuples $(N_\uparrow, N_\downarrow, \nu)$ marked with an asterisk (*) indicate an open shell at the Fermi level. Uncertainties smaller than $5 \times 10^{-5}$ ($5 \times 10^{-3}$) are reported as (0) for energies (relative errors).
    }
    \centering
    \begin{tabular}{cccccccc}
        \toprule
        \multirow{2}{*}{$(N_\uparrow, N_\downarrow, \nu)$} & \multirow{2}{*}{$U$} & Ground state & \multirow{2}{*}{Exact} & CPMC & Rel. & CPMC & Rel. \\
        & & symmetry & & (FE) & error (\%) & [($K$, SG, $S^2$)-GHF] & error (\%) \\
        \midrule
        \multirow{3}{*}{(2, 2, 1/4)} & 4 & $A_1$ & $-1.0746$ & $-1.0746(0)$ & 0.00(0) & $-1.0746(0)$ & 0.00(0) \\
                                     & 8 & $A_1$ & $-1.0483$ & $-1.0483(0)$ & 0.00(0) & $-1.0483(0)$ & 0.00(0) \\
                                     & 12 & $A_1$ & $-1.0320$ & $-1.0320(0)$ & 0.00(0) & $-1.0322(1)$ & $-0.02(0)$ \\
        \midrule
        \multirow{3}{*}{$(3, 3, 3/8)^*$} & 4 & $B_1$ & $-1.2296$ & $-1.2298(0)$ & $-0.01(0)$ & $-1.2298(0)$ & $-0.01(0)$ \\
                        & 8 & $B_1$ & $-1.1779$ & $-1.1787(1)$ & $-0.06(0)$ & $-1.1771(2)$ & 0.07(2) \\
                        & 12 & $B_1$ & $-1.1467$ & $-1.1477(1)$ & $-0.09(1)$ & $-1.1467(1)$ & 0.00(1) \\
        \midrule
        \multirow{3}{*}{(4, 4, 1/2)} & 4 & $A_1$ & $-1.3408$ & $-1.3409(0)$ & $-0.01(0)$ & $-1.3406(1)$ & 0.01(1) \\
                    & 8 & $B_1$ & $-1.2520$ & $-$ & $-$ & $-1.2517(1)$ & 0.02(0) \\
                    & 12 & $B_1$ & $-1.2041$ & $-$ & $-$ & $-1.2034(1)$ & 0.06(1) \\
        \midrule
        \multirow{3}{*}{$(6, 6, 3/4)^*$} & 4 & $B_1$ & $-1.3703$ & $-1.3706(1)$ & $-0.03(0)$ & $-1.3704(1)$ & $-0.01(1)$ \\
                        & 8 & $B_1$ & $-1.1648$ & $-1.1670(2)$ & $-0.19(2)$ & $-1.1634(2)$ & 0.12(1) \\
                        & 12 & $B_1$ & $-1.0569$ & $-1.0603(3)$ & $-0.32(3)$ & $-1.0517(3)$ & 0.49(2) \\
        \midrule
        \multirow{3}{*}{(7, 7, 7/8)} & 4 & $A_1$ & $-1.2612$ & $-1.2611(1)$ & 0.01(1) & $-1.2614(1)$ & $-0.02(1)$ \\
                        & 8 & $A_1$ & $-0.9547$ & $-0.9558(3)$ & $-0.11(3)$ & $-0.9530(1)$ & 0.18(1) \\
                        & 12 & $A_1$ & $-0.8079$ & $-0.8086(5)$ & $-0.09(6)$ & $-0.8023(2)$ & 0.69(2) \\
        \midrule
        \multirow{3}{*}{$(8, 8, 1)^*$} & 4 & $B_1$ & $-0.9725$ & $-0.9739(1)$ & $-0.14(1) $& $-0.9724(1)$ & $0.02(1)$ \\
                      & 8  & $B_1$ & $-0.5294$ & $-0.5343(5)$ & $-0.93(9)$ & $-0.5239(1)$ & $1.03(2)$ \\
                      & 12 & $A_1$ & $-0.3611$ & $-0.3394(6)$ & 6.01(17) & $-0.3584(2)$ & $0.75(5)$ \\
        \bottomrule
    \end{tabular}
    \label{tab:doped}
\end{table*}

In this work, we consider generalized Hartree-Fock (GHF) determinants that break both $\hat{S}_z$ and $\hat{S}^2$ spin symmetries, enabling greater variational freedom in describing the frustration inherent in the triangular-lattice Hubbard model. Furthermore, symmetry-restored GHF trial states have also been shown to yield better accuracy compared to unrestricted Hartree-Fock (UHF) trial states that only break $\hat{S}_z$ symmetry \cite{shi_symmetry-projected_2014}. We variationally optimize the GHF state orbital coefficients in the presence of symmetry-projection operators (\textit{i.e.} the variation-after-projection (VAP) approach), leading to a lower energy trial state (see App.~\ref{app:symm_trial} for details). We follow Ref.~\cite{Jimenez-Hoyos2012} in using the notation $x$-GHF to denote symmetry-projected wave functions, where $x$ indicates the symmetries broken and restored. For example, ($K$, SG, $S^2$)-GHF refers to a state prepared by breaking and restoring spin ($S^2$), space-group (SG), and complex conjugation ($K$) symmetry out of a GHF determinant.

We also consider a class of symmetry-preserving FE trials consisting of Slater determinants constructed from the one-body eigenstates of $\hat{H}_1$. A single determinant often preserves the Hamiltonian symmetries for closed-shell systems, but multiple determinants are required for open-shell systems to preserve spin and spatial symmetries. We distinguish between these cases using c-FE (for closed-shell) and o-FE (for open-shell).

\section{Results} \label{results}
\subsection{Away from half-filling}

\begin{figure*}[t]
    \centering
    \includegraphics[width=\linewidth]{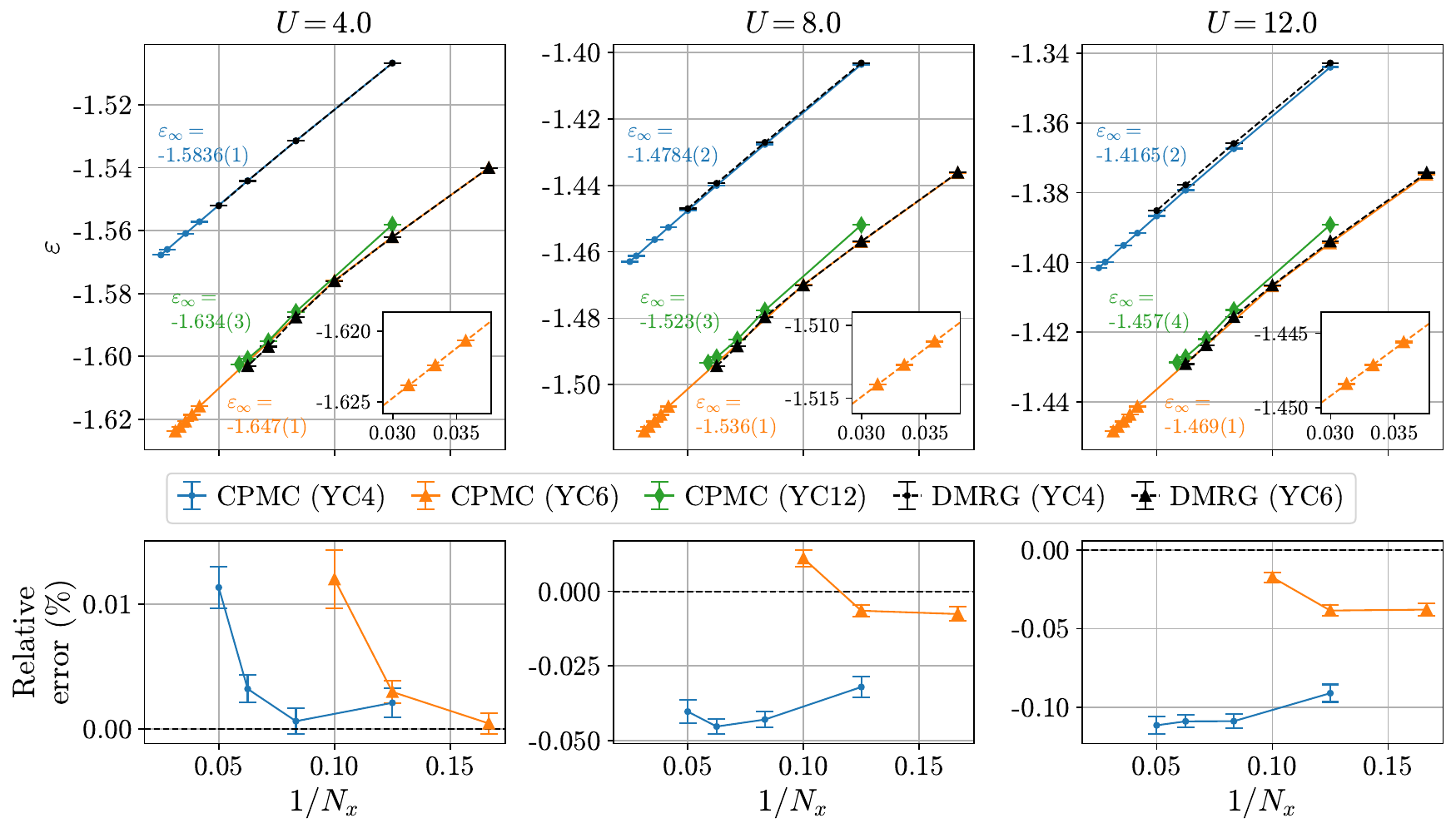}
    \caption{(Top panels) Ground state energies per site obtained from CPMC and DMRG on YC$n$ cylinders with widths $n \equiv N_y = 4, 6, 12$ at quarter-filling ($\nu = 1/2$) and $U = 4, 8, 12$. The insets highlight the three largest $N_x$ points, which exhibit a clear linear dependence on $N_x^{-1}$. Colored labels indicate the extrapolated infinite-cylinder energies $\varepsilon_{\infty}$ obtained from data of the corresponding color. The quoted uncertainties include both statistical and systematic contributions from the fitting procedure, detailed in App.~\ref{app:cpmc_extrap}. (Bottom panels) Relative error between CPMC and DMRG energies for the YC4 and YC6 cylinders, which remain below $1\%$.}
    \label{fig:doped_energies}
\end{figure*}

Table \ref{tab:doped} compares the ground state energies per site $\varepsilon$ obtained from CPMC and ED for the XC $4 \times 4$ lattice as an illustrative example; data for other lattices can be found in Table \ref{tab:more_data} in the appendix. We find that CPMC is highly accurate away from half-filling, achieving relative errors $<0.5\%$ from ED over a range of $U$. This is attained using relatively simple trial wave functions that preserve the symmetry of the ground state. 

In all cases considered here, the ground state is non-degenerate with spin $s = 0$ and momentum (0, 0). At fillings where the system is closed-shell ($\nu = 1/4, 1/2, 7/8$), the ground state generally transforms as the $A_1$ irrep of the lattice symmetry group $D_4$, hence c-FE trials were used since they reside in the same symmetry sector. For $U > 4$ at $\nu = 1/2$, however, the ground state symmetry transitions from $A_1$ to $B_1$, so the c-FE trial is no longer adequate since it has zero overlap with the ground state. We thus only report CPMC energies using symmetry-projected trials. The effects of symmetry projection and its necessity at half-filling are illustrated in further detail in Section \ref{half_filling}. 

At fillings where the system is open-shell ($\nu = 3/8, 3/4$), the ground state transforms as the $B_1$ irrep of $D_4$. We prepare the symmetry-adapted o-FE trial by first choosing a $D_2$-symmetry-adapted basis within the degenerate subspace, \textit{i.e.} forming linear combinations of the orbitals that transform as irreps of a $D_2 \subset D_4$ subgroup (see Fig.~\ref{fig:lattices}). We then define a complete active space (CAS) expansion in this subspace that gives a state with the target $D_4$ symmetry. To be explicit, the o-FE trial for $\nu = 3/8 \ (N_\uparrow = N_\downarrow = 3)$ is given by 
\begin{align}
    \ket{\psi_T} &= \frac{1}{\sqrt{2}} \ket{A_1^{(1)} \bar{A}_1^{(1)} A_2^{(1)} \bar{A}_2^{(1)} B_1^{(1)} \bar{B}_1^{(1)}} \nonumber \\
    & \qquad - \frac{1}{\sqrt{2}}  \ket{A_1^{(1)} \bar{A}_1^{(1)} A_2^{(1)} \bar{A}_2^{(1)} B_2^{(1)} \bar{B}_2^{(1)}}, \label{doped_fe_trial}
\end{align}
where $\Gamma^{(n)} \ (\bar{\Gamma}^{(n)})$ denotes the $n$-th level up- (down-) spin orbitals in irrep $\Gamma$ of the $D_2$ subgroup (see Fig. \ref{fig:lattices}) and each ket lists the occupied spin orbitals defining a Slater determinant. App.~\ref{app:xc4_symm} contains further details on symmetry characterization.

Having performed a similar study on small YC$n$ lattices and obtained consistent results there (see Table \ref{tab:more_data} in App.~\ref{app:more_data}), we next benchmark CPMC against DMRG on long YC$n$ cylinders at quarter-filling ($\nu = 1/2$). This choice is motivated by two observations: (i) the system is often closed-shell for such cylinders at this filling; and (ii) the ground state of the YC $4 \times 4$ and $4 \times 3$ lattices at $\nu = 1/2$ transforms as the $A_1$ irrep of $D_4$ for all $U$ considered, suggesting that this behavior may generalize to wider YC$n$ cylinders. Thus, we restrict our attention to closed-shell systems and employ c-FE trial wave functions here. 

As shown in Fig. \ref{fig:doped_energies}, CPMC achieves comparable accuracy to DMRG and exhibits a linear relationship with $N_x^{-1}$ for a fixed $N_y$. This arises from the presence of open boundaries along the $x$-direction, which causes the energy per site to approach the infinite cylinder limit $N_x \to \infty$ with an error proportional to $N_x^{-1}$ \cite{stoudenmire_studying_2012}. In contrast to DMRG's steep scaling with the width $N_y$, the cost of CPMC only scales cubically with the number of sites $N_s$, enabling CPMC to reach system sizes that are inaccessible to DMRG. We demonstrate this on YC$n$ cylinders up to $n = N_y = 12$ and provide estimates of infinite cylinder bulk energies in Fig. \ref{fig:doped_energies} by extrapolating in $N_x^{-1}$ (see App.~\ref{app:cpmc_extrap} for details). Thus, once infinite cylinder bulk energies are obtained, it is straightforward to extrapolate in $N_y$ to estimate the 2D bulk energy.

\subsection{Half-filling} \label{half_filling}
 
\begin{figure*}[t]
    \centering
    \includegraphics[width=\linewidth]{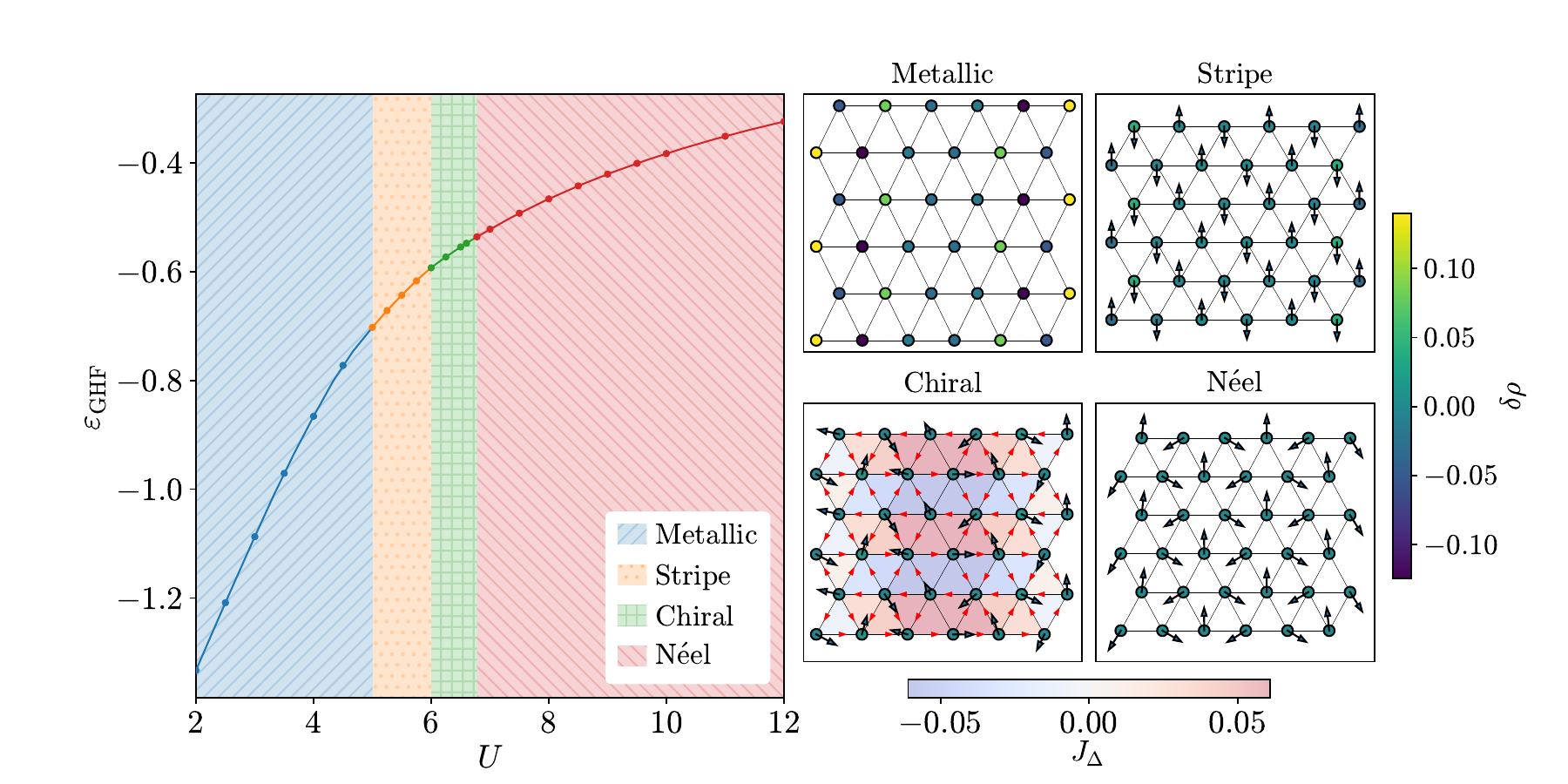}
    \caption{
        GHF ground state solutions on XC$n$ cylinders at half-filling as a function of $U$. We find a metallic state at $U \lesssim 5$ and a Néel state at $U \gtrsim 6.8$. At intermediate $U$, we identify a stripe state at $5 \lesssim U \lesssim 6$ and a chiral state at $6 \lesssim U \lesssim 6.8$. The right panels depict the local spins and density fluctuations $\delta \rho = \rho - \braket{\rho}$ exhibited by each state. For the chiral state, we also show the bond charge current $J_{ij}$ and plaquette current $J_\triangle$, defined in \eqref{bond_charge_current}-\eqref{plaquette_current}.
    }
    \label{fig:xc_6x6_ghf}
\end{figure*}

In contrast to the square lattice, the triangular lattice is not bipartite, so particle-hole symmetry is absent in the Hubbard model at all fillings. This implies that the sign problem persists even at half-filling ($\nu = 1$), where we find the associated constraint bias more challenging to mitigate than in the doped cases. In particular, we achieve $\lesssim 1\%$ relative errors at large $U$ only when employing symmetry-projected trial wave functions that yield the lowest variational energies while preserving the essential symmetries of the system. 

The GHF phase diagram for XC$n$ cylinders is shown in Fig. \ref{fig:xc_6x6_ghf}, displaying four distinct solutions as a function of $U$. The ground state transitions from a metallic phase in the weak-coupling regime $U \lesssim 4$ to a vertical stripe phase with alternating spin orders along the $x$-direction at intermediate $U$. At $U \approx 6$, the lowest GHF solution breaks complex conjugation symmetry and exhibits a chiral phase with non-zero bond charge currents \cite{zhu_doped_2022}
\begin{align}
    J_{ij} = -it \sum_{\sigma} \left[\braket{\hat{c}^\dagger_{i\sigma} \hat{c}_{j\sigma}} - \braket{\hat{c}^\dagger_{j\sigma} \hat{c}_{i\sigma}}\right] \label{bond_charge_current}
\end{align}
between nearest-neighbor sites $i, j$. The emergence of chiral domains is seen more clearly by computing the plaquette current on an elementary triangular plaquette $(i, j, k)$, defined as
\begin{align}
    J_{\triangle, ijk} = J_{ij} + J_{jk} + J_{ki}, \label{plaquette_current}
\end{align}
where the loop direction is taken counter-clockwise (see Fig. \ref{fig:xc_6x6_ghf}). This state lies close to the $120^\circ$ Néel-ordered phase that persists into the strong-coupling regime $U \gtrsim 10$ and may be related to the spiral magnetic phases reported in previous GHF studies \cite{jayaprakash_metal-insulator_1991,gilmutdinov_v_f_spiral_2022}. We note that GHF overestimates the stability of the insulating phases, predicting their onset at $U \approx 5$ in contrast to DMRG's prediction of $U \approx 9$ \cite{szasz_chiral_2020,szasz_phase_2021,chen_quantum_2022}.

From these GHF states, we prepare symmetry-projected trials by restoring spin, spatial, and complex conjugation symmetries via a variation-after-projection approach. Since the half-filled system is open-shell, another reasonable trial wave function is a symmetry-adapted o-FE trial constructed to match the $D_4$ irrep of the exact ground state (which varies with $U$; see Table \ref{tab:doped}). For $U < 12$, the ground state transforms as the $B_1$ irrep of $D_4$, so in the same notation as \eqref{doped_fe_trial}, we define (see Fig.~\ref{fig:lattices})
\begin{align}
    \ket{\psi_T} &= \frac{1}{\sqrt{2}} \ket{A_1^{(1)} \cdots B_1^{(3)} \bar{B}_1^{(3)}} - \frac{1}{\sqrt{2}} \ket{A_1^{(1)} \cdots B_2^{(3)} \bar{B}_2^{(3)}}. \label{half_fe_trial_b1}
\end{align}
The orbitals are labeled by $D_2$ irreps, and we omit orbitals below the Fermi level for clarity. For $U = 12$, the ground state transforms as the $A_1$ irrep of $D_4$, and we have 
\begin{align}
    \ket{\psi_T} &= \frac{1}{\sqrt{2}} \ket{A_1^{(1)} \cdots B_1^{(3)} \bar{B}_1^{(3)}} + \frac{1}{\sqrt{2}} \ket{A_1^{(1)} \cdots B_2^{(3)} \bar{B}_2^{(3)}}. \label{half_fe_trial_a1}
\end{align}

\begin{figure}[t]
    \centering
    \includegraphics[width=\linewidth]{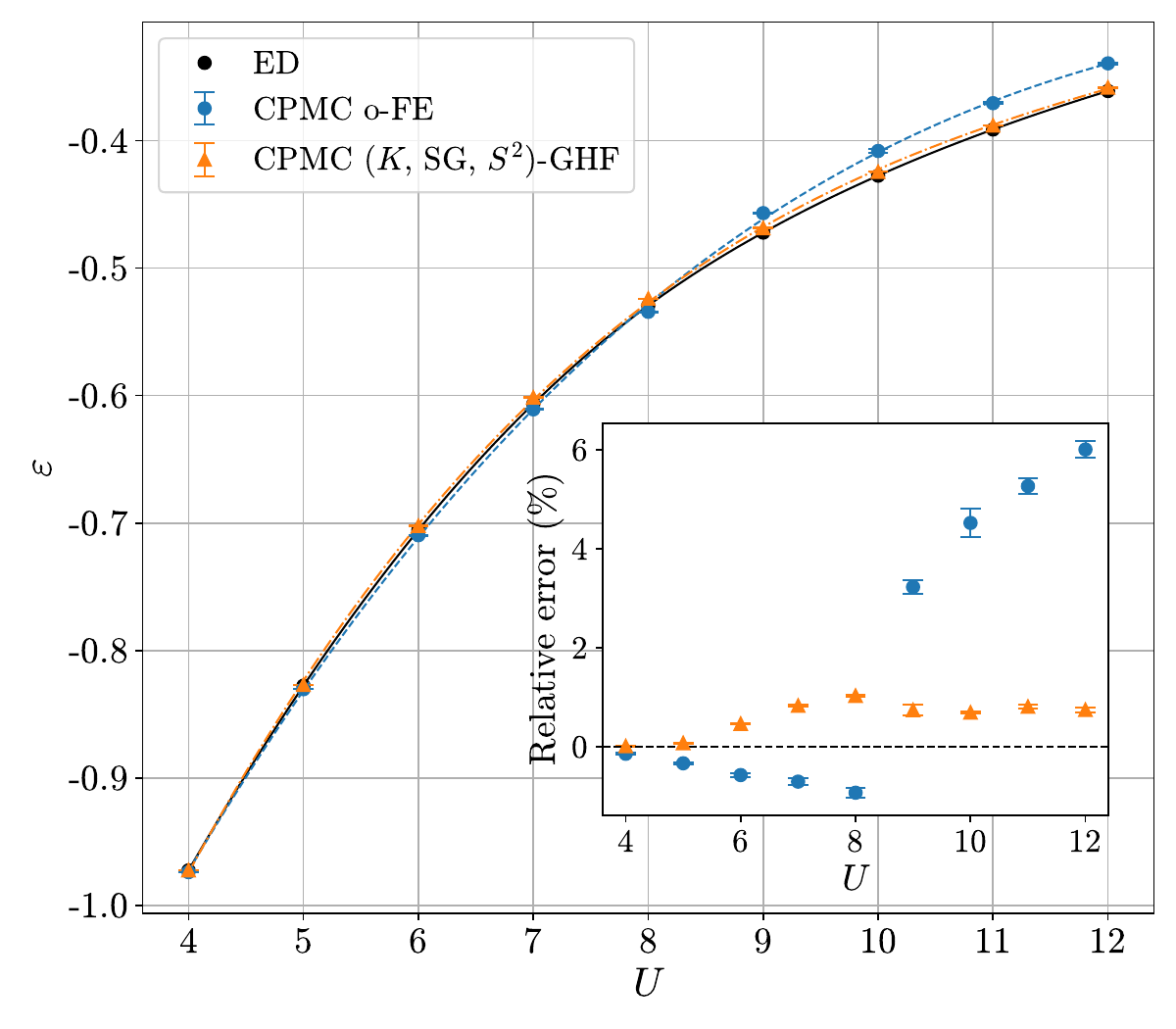}
    \caption{Ground state energies per site obtained from CPMC and ED for the XC $4 \times 4$ lattice at half-filling ($\nu = 1$). The inset depicts the relative error between CPMC and ED energies. Employing the symmetry-projected ($K$, SG, $S^2$)-GHF trial yields improved energies compared to the o-FE trial.}
    \label{fig:xc_4x4_cpmc_scan_U}
\end{figure}

\begin{figure}[t]
    \centering
    \includegraphics[width=\linewidth]{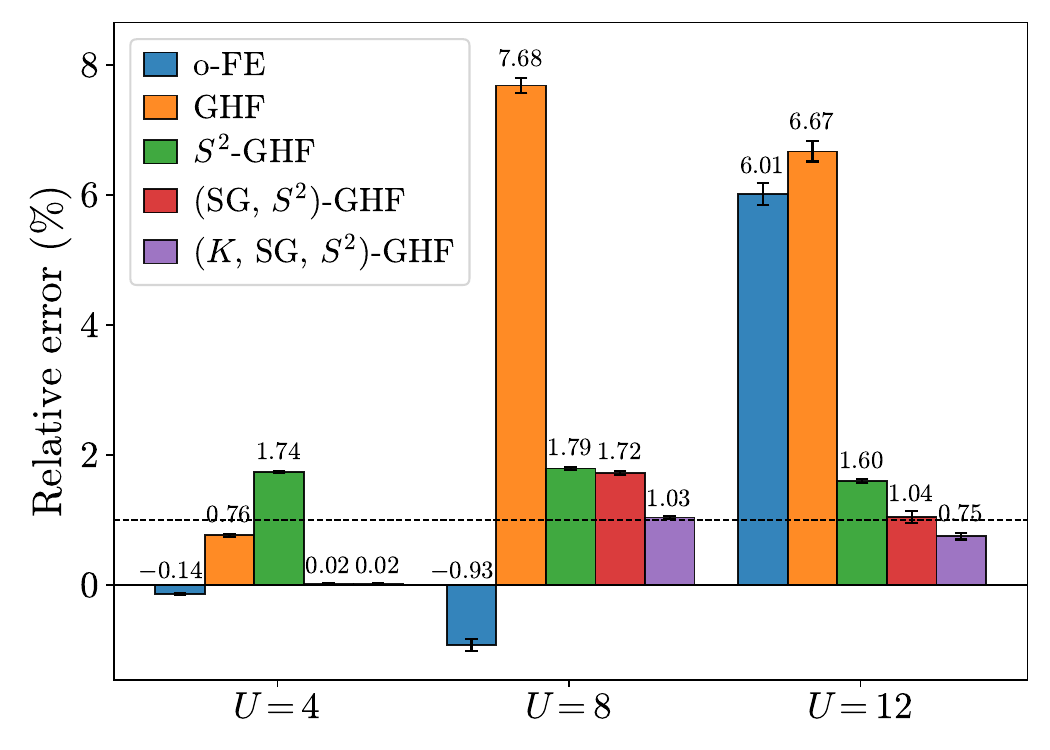}
    \caption{Relative error between CPMC and ED energies for the XC $4 \times 4$ lattice at $U = 4, 8, 12$ from o-FE and a variety of symmetry-projected trials. The relative errors are systematically reduced as symmetry projectors are successively applied onto the GHF state.}
    \label{fig:xc_4x4_symm_projection}
\end{figure}

A comparison of CPMC energies with o-FE and ($K$, SG, $S^2$)-GHF trial wave functions against ED for the XC $4 \times 4$ lattice is shown in Fig. \ref{fig:xc_4x4_cpmc_scan_U}. Substantial improvement is observed with ($K$, SG, $S^2$)-GHF trials--especially for $U \gtrsim 9$--where relative errors are $\lesssim 1\%$ compared to $\sim 3-6\%$ obtained with o-FE trials, which increase monotonically with $U$. This illustrates the significance of these symmetries ($K$, SG, $S^2$) and the inadequacy of FE trials in describing the ground state as the system becomes more strongly correlated. Fig. \ref{fig:xc_4x4_symm_projection} shows the improvement in the trial quality as $S^2$, SG, and $K$ symmetry projectors are successively applied. Across the range of $U$ values considered, deviations from ED are systematically reduced as more symmetries are restored from the GHF state. 

The importance of each symmetry can be inferred from the reduction in error as the corresponding symmetry projector is applied. The broken-symmetry GHF trial yields the largest relative errors, except at $U = 4$. In this regime, restoring both spatial and spin symmetries is necessary to improve upon the GHF trial, while additionally restoring complex conjugation symmetry has a negligible effect. By contrast, restoring spin symmetry alone yields a larger relative error than GHF, highlighting the coupled role of spin and spatial symmetries at low $U$.

At $U = 8$, restoring spin symmetry yields a substantial reduction in relative error compared to GHF, whereas further restoring spatial symmetry provides only a marginal improvement. Including complex conjugation symmetry gives the lowest error, suggesting that spin and complex conjugation symmetries are most important for describing the ground state in this regime. We note that the projected-energy landscape, $\braket{\Psi_{\text{GHF}} | \hat{H} \hat{\mathcal{P}} | \Psi_{\text{GHF}}}$, over the determinants $\ket{\Psi_{\text{GHF}}}$ exhibits several local minima, making the global minimum difficult to locate. We select the trial wave function based on two criteria: (i) the optimization converges to $10^{-5}$ in the gradient norm; and (ii) it has the lowest variational energy among the solutions found. The lowest-variational-energy trial we found yields a CPMC energy with a relative error of 1.03\%.

At $U = 12$, the relative error decreases monotonically as more symmetries are restored on top of the preceding trial. Specifically, the error is reduced by factors of $\sim 4.2$ from GHF $\rightarrow$ $S^2$-GHF, $\sim 1.5$ from $S^2$-GHF $\rightarrow$ (SG, $S^2$)-GHF, and $\sim 1.4$ from (SG, $S^2$)-GHF $\rightarrow$ ($K$, SG, $S^2$)-GHF, again suggesting that spin symmetry is most important, followed by spatial and complex conjugation symmetries. 

Another observation worth pointing out in Fig. \ref{fig:xc_4x4_symm_projection} is the efficacy of o-FE trials compared to GHF. To rationalize this, recall from \eqref{estimator} that the CPMC energy $E_\text{CPMC}$ is estimated from the weights $w_i$ and local energies $E_L[\phi_i]$,
\begin{align}
    E_\text{CPMC} \equiv \braket{\hat{H}} = \frac{\sum_i w_i E_L[\phi_i]}{\sum_i w_i}, \quad E_L[\phi_i] = \frac{\braket{\psi_T | \hat{H} | \phi_i}}{\braket{\psi_T | \phi_i}}, \label{energy_estimator}
\end{align}
where $\ket{\psi_T}$ is the trial wave function. One can see that $\ket{\psi_T}$ affects $E_\text{CPMC}$ through: (i) importance sampling and applying the constraint in $w_i$ and $\ket{\phi_i}$; and (ii) the mixed estimator $E_L$. Although the o-FE wave function has a high variational energy, which may incur a large bias through $E_L$ in \eqref{energy_estimator}, it preserves the expected symmetries of the ground state and therefore restricts the constrained random walk to a more appropriate sector in Slater determinant space. This symmetry consistency may explain why o-FE trials yield smaller errors compared to GHF trials, in line with the improvements observed for symmetry-projected trials. On the other hand, the increasing error of o-FE trials with $U$ is likely due to the growing mismatch between the o-FE wave function and the strongly-correlated ground state.

\begin{figure*}[ht]
    \centering
    \includegraphics[width=\linewidth]{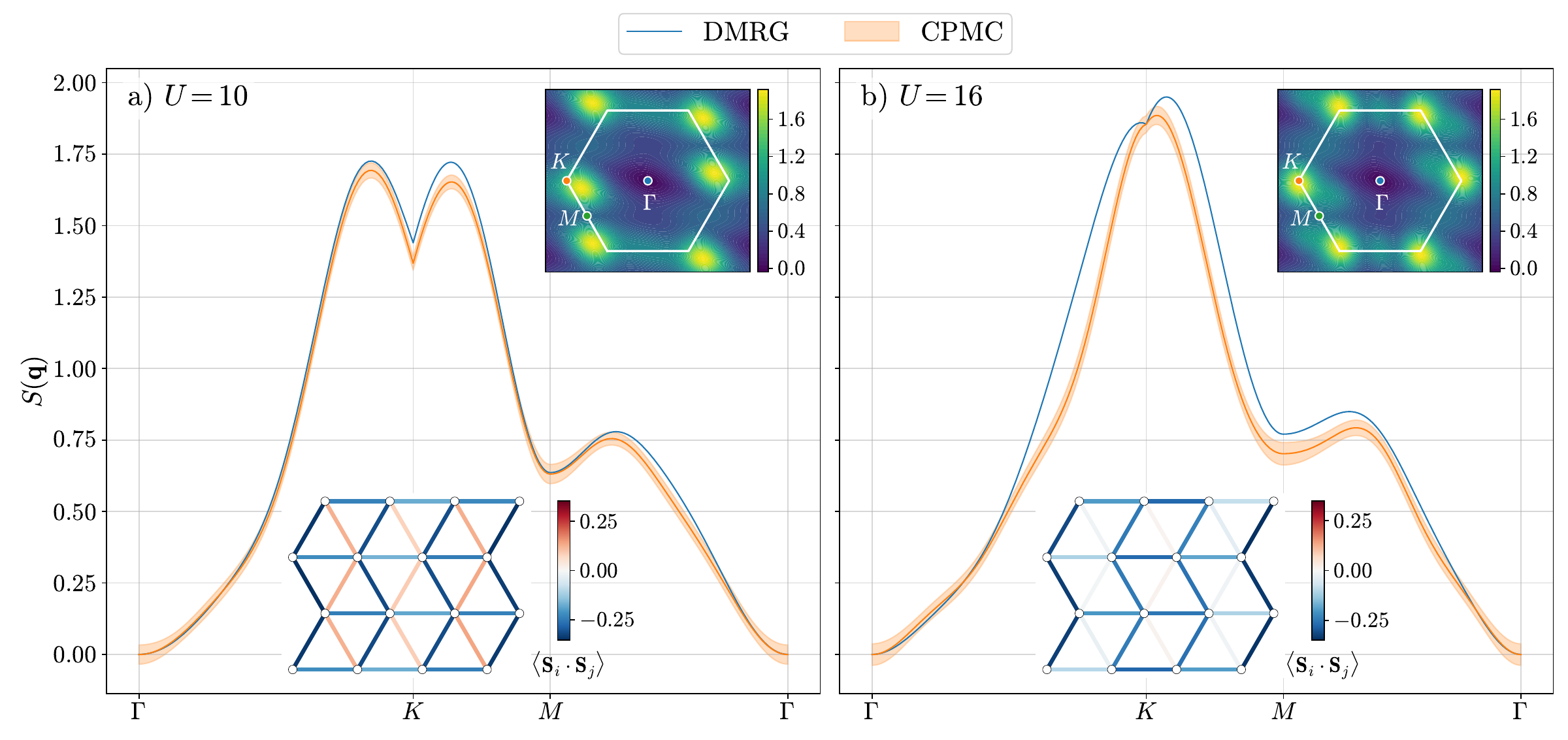}
    \caption{Spin structure factor $S(\mathbf{q})$ for the XC $4\times4$ lattice at $U=10$ (left panel) and $U=16$ (right panel). In each panel, the main plot compares CPMC and DMRG results for $S(\mathbf{q})$ along the $\Gamma-K-M-\Gamma$ high-symmetry path in the Brillouin zone. The insets show $S(\mathbf{q})$ plotted over the Brillouin zone (top right) and the real-space spin correlations $\braket{\hat{\mathbf{S}}_i \cdot \hat{\mathbf{S}}_j}_c$ (bottom) obtained using CPMC.}
    \label{fig:ssf}
\end{figure*}

In addition to the energy, we also evaluate the spin structure factor,
\begin{equation}
   S(\mathbf{q}) = \frac{1}{N_s}\sum_{ij}e^{i\mathbf{q}\cdot(\mathbf{r}_i - \mathbf{r}_j)} \langle \hat{\mathbf{S}}_i \cdot \hat{\mathbf{S}}_j \rangle_c,
\end{equation}
where $\langle \hat{\mathbf{S}}_i \cdot \hat{\mathbf{S}}_j \rangle_c$ is the \emph{connected} spin-spin correlation function, $\hat{\mathbf{S}}_i = (\hat{S}_i^x, \hat{S}_i^y, \hat{S}_i^z)$ is the spin operator at site $i$, and $\mathbf{r}_i$ is its position vector. We estimate $S(\mathbf{q})$ within CPMC using the mixed estimator and compare with SU(2) spin-symmetry-preserving DMRG for the XC $4 \times 4$ lattice at $U = 10$ and $U = 16$ (Fig.~\ref{fig:ssf}). The $(K,\text{SG},S^2)$-GHF trial wave function is used. For both interaction strengths, we find good agreement between CPMC and DMRG. 

At $U = 10$, stripy correlations are already apparent in this small lattice, with ferromagnetic correlations along one direction and antiferromagnetic correlations along the other two, consistent with previous studies \cite{wietek_mott_2021,chen_quantum_2022}. Upon increasing $U$ to 16, the stripy correlations evolve into the tripartite $120^\circ$ Néel order. For larger widths, it becomes challenging to converge the DMRG spin structure factor without breaking symmetries, likely due to the one-dimensional site ordering used in DMRG. A systematic study of the evolution of spin correlations with system size is left for future work.

\section{Discussion} \label{discussion}
In this work, we present a benchmark study of the triangular-lattice Hubbard model at several fillings and $U$ values using constrained-path Monte Carlo. Our findings show that the use of appropriate symmetry-adapted trial wave functions is important for obtaining accurate results. Away from half-filling, we achieve energy deviations of $\lesssim 0.5\%$ compared to ED (on small clusters) and DMRG (on long cylinders) using relatively simple FE trial wave functions that preserve the symmetry of the ground state. At half-filling, we first identify four distinct GHF solutions--metallic, stripe, chiral, and Néel--as a function of $U$. We then highlight the inadequacy of using these broken-symmetry states as trial wave functions and the necessity of symmetry restoration in the intermediate to large $U$ regimes. Here, unprojected GHF trials can introduce a sizable constraint bias that is mitigated by restoring the relevant symmetries. This should be borne in mind when interpreting conclusions drawn from CPMC studies of the half-filled and doped triangular-lattice Hubbard models, including the results of Ref. \cite{zampronio_chiral_2023}. Therein, the authors report evidence of chiral superconductivity using trial wave functions prepared from GHF states computed at an effective $U$ value; however, symmetry considerations do not appear to factor into trial construction. We demonstrate that hierarchically breaking then restoring spin, spatial, and complex conjugation symmetries can systematically increase the quality of the trial wave function for CPMC, providing a framework for assessing such trial-state descriptions of this frustrated system.

Our study is limited to cylinder geometries to facilitate comparison with DMRG results and mitigate open-shell effects that arise from degeneracies in the one-electron spectrum. The latter can be alleviated by employing twist boundary conditions that lift these degeneracies; averaging over random twists (\textit{i.e.} twist-averaged boundary conditions) may also accelerate convergence to the thermodynamic limit \cite{qin_benchmark_2016}. We leave the investigation of the performance of CPMC under such boundary conditions to future work. 

Finally, we comment on size extensivity. Although symmetry-projected Hartree-Fock wave functions are not size extensive \cite{Jimenez-Hoyos2012,lee_twenty_2022}, previous work suggests that CPMC with symmetry-projected trials can nevertheless exhibit well-behaved finite-size scaling \cite{shi_symmetry-projected_2014}. In the square-lattice Hubbard model with periodic boundary conditions, CPMC energies obtained using $S^2$-projected UHF trials were shown to approach estimates of the exact ground-state energy in the thermodynamic limit with increasing system sizes up to $14 \times 14$, albeit non-monotonically \cite{shi_symmetry-projected_2014}. Whether analogous behavior occurs for symmetry-projected trials on the triangular lattice remains an open question and is also left for future study.

The existence of a spin-liquid state in the triangular-lattice Hubbard model remains a controversial topic, with various numerical techniques providing evidence for both sides of the debate. On the one hand, DMRG studies on long cylinders generally suggest the presence of a spin-liquid phase, while variational Monte Carlo studies do not. This discrepancy could perhaps be attributed to cylindrical geometries that bias the ground state towards the spin liquid. The non-trivial effects of the cylinder length have been highlighted in \cite{sandvik_finite-size_2012} and \cite{chen_quantum_2022}, the latter of which reports that long-range chiral correlations are only observed in sufficiently long cylinders. While DMRG's exponential scaling in the system's width limits its applicability to cylinders, the computational cost of CPMC scales only polynomially with the system size, enabling an approach to the thermodynamic limit via two-dimensional sheets of increasing area. Therefore, CPMC may be a valuable tool for investigating the existence of spin liquids in the triangular-lattice Hubbard model. 

\section{Acknowledgments}
We acknowledge funding support from the National Science Foundation (NSF) through the Columbia University Materials Research Science and Engineering Center (MRSEC) on Precision Assembled Quantum Materials under Award No. DMR-2011738 (S.F.U. and D.R.R.) and the Air Force Office of Scientific Research (AFOSR) under MURI Grant No. FA 9550-25-1-0288 (A.M. and D.R.R.). We thank Shiwei Zhang, Joonho Lee, Tong Jiang, Moritz Baumgarten, and Pavan Ravindra for useful discussions. Computations in this paper were primarily executed on the Delta system at the National Center for Supercomputing Applications [award OAC 2005572] through allocation CHE230028 from the Advanced Cyberinfrastructure Coordination Ecosystem: Services \& Support (ACCESS) program, which is supported by National Science Foundation grants \#2138259, \#2138286, \#2138307, \#2137603, and \#2138296. We also acknowledge computing resources from Columbia University's Shared Research Computing Facility.

\section{Data availability}
Scripts and data used to generate the figures is available at \cite{data}.

\clearpage

\appendix
\onecolumngrid
\section{Additional CPMC data and computational details}
\subsection{Additional data} \label{app:more_data}
Table~\ref{tab:more_data} contains additional CPMC data for XC $3 \times 4$, YC $4 \times 3$, and YC $4 \times 4$ cylinders at several fillings and $U$ values. They again show that CPMC with a simple FE trial is highly accurate away from half filling, but symmetry projection may be important for achieving similar accuracy closer to half filling.

\begin{table*}[h!]
    \setlength{\belowcaptionskip}{10pt}
    \caption{
        Ground state energies per site obtained from CPMC and ED for fillings $\nu \leq 1$. The relative (rel.) error is computed as $(\varepsilon_\text{CPMC} - \varepsilon_\text{exact}) / |\varepsilon_\text{exact}| \times 100$. Tuples $(N_\uparrow, N_\downarrow, \nu)$ marked with an asterisk (*) indicate an open shell at the Fermi level. CPMC energies marked with a dagger $(\dagger)$ are obtained with symmetry projection; otherwise, FE trials are used. Uncertainties smaller than $5 \times 10^{-5}$ ($5 \times 10^{-3}$) are reported as (0) for energies (relative errors).
    }
    \centering
    \begin{tabular}{ccccccc}
        \toprule
        Lattice ($N_x \times N_y$) & $(N_\uparrow, N_\downarrow, \nu)$ & $U$ & Ground state symmetry & CPMC & Exact & Rel. error (\%) \\
        \midrule
        \multirow{10}{*}{XC $3 \times 4$} & \multirow{2}{*}{(2, 2, 1/3)} & 6 & $A_1$ & $-1.1376(1)$ & $-1.1375$ & $-0.01(1)$ \\
                     &             & 12 & $A_1$ & $-1.0981(1)$ & $-1.0978$ & $-0.02(1)$ \\
        \cmidrule{2-7}
                     & \multirow{2}{*}{$(3, 3, 1/2)^*$} & 6 & $B_1$ & $-1.2649(1)$ & $-1.2675$ & 0.21(1) \\
                     &             & 12 & $B_1$ & $-1.1853(2)$ & $-1.1843$ & $-0.08(2)$ \\
        \cmidrule{2-7}
                     & \multirow{2}{*}{(4, 4, 2/3)} & 6 & $A_1$ & $-1.2868(2)$ & $-1.2865$ & $-0.03(2)$ \\
                     &             & 12 & $A_1$ & $-1.1350(2)$ & $-1.1337$ & $-0.11(2)$ \\
        \cmidrule{2-7}
                     & \multirow{2}{*}{$(5, 5, 5/6)^*$} & 6 & $B_1$ & $-1.0871(2)$ & $-1.0859$ & $-0.11(2)$ \\
                     &             & 12 & $B_1$ & $-0.8709(4)$ & $-0.8689$ & $-0.23(5)$ \\
        \cmidrule{2-7}
                     & \multirow{2}{*}{(6, 6, 1)} & 6  & $B_1$ & $-0.6584(2)^\dagger$ & $-0.6600$ & $0.23(3)$ \\
                     &             & 12 & $B_1$ & $-0.3435(2)^\dagger$ & $-0.3465$ & 0.86(5) \\
        \midrule
        \multirow{8}{*}{YC $4 \times 3$} & \multirow{2}{*}{(2, 2, 1/3)} & 6 & $A_1$ & $-1.3045(0)$ & $-1.3045$ & 0.00(0) \\
                     &             & 12 & $A_1$ & $-1.2562(0)$ & $-1.2561$ & $-0.01(0)$ \\
        \cmidrule{2-7}
                     & \multirow{2}{*}{(3, 3, 1/2)} & 6 & $A_1$ & $-1.3183(0)$ & $-1.3189$ & 0.04(0) \\
                     &             & 12 & $A_1$ & $-1.2298(2)$ & $-1.2313$ & 0.12(2) \\
        \cmidrule{2-7}
                     & \multirow{2}{*}{$(4, 4, 2/3)^*$} & 6 & $\prescript{3}{}{A_2}$ & $-1.2615(2)$ & $-1.2610$ & $-0.04(1)$ \\
                     &             & 12 & $\prescript{3}{}{A_2}$ & $-1.1163(3)$ & $-1.1170$ & 0.07(2) \\
        \cmidrule{2-7}
                     & \multirow{2}{*}{(5, 5, 5/6)} & 6 & $B_1$ & $-1.0740(1)^\dagger$ & $-1.0741$ & $0.01(1)$ \\        
                     &             & 12 & $B_1$ & $-0.8716(2)^\dagger$ & $-0.8748$ & 0.36(2) \\
        \midrule
        \multirow{15}{*}{YC $4 \times 4$} & \multirow{3}{*}{(2, 2, 1/4)} & 4 & $A_1$ & $-1.0157(0)$ & $-1.0157$ & 0.00(0) \\
                     &             & 8 & $A_1$ & $-0.9934(0)$ & $-0.9934$ & 0.00(0) \\
                     &             & 12 & $A_1$ & $-0.9801(0)$ & $-0.9800$ & -0.01(0) \\
        \cmidrule{2-7}
                     & \multirow{3}{*}{$(3, 3, 3/8)^*$} & 4 & $\prescript{3}{}{A_2}$ & $-1.2530(0)$ & $-1.2529$ & 0.00(0) \\
                     &             & 8 & $\prescript{3}{}{A_2}$ & $-1.2028(0)$ & $-1.2023$ & $-0.04(0)$ \\
                     &             & 12 & $\prescript{3}{}{A_2}$ & $-1.1730(1)$ & $-1.1720$ & $-0.08(1)$ \\
        \cmidrule{2-7}
                     & \multirow{3}{*}{(4, 4, 1/2)} & 4 & $A_1$ & $-1.4382(0)$ & $-1.4383$ & 0.00(0) \\
                     &             & 8 & $A_1$ & $-1.3371(1)$ & $-1.3366$ & $-0.04(0)$ \\
                     &             & 12 & $A_1$ & $-1.2796(1)$ & $-1.2785$ & $-0.08(1)$ \\
        \cmidrule{2-7}
                     & \multirow{3}{*}{(6, 6, 3/4)} & 4 & $A_1$ & $-1.4382(1)$ & $-1.4387$ & 0.03(1) \\
                     &             & 8 & $A_1$ & $-1.2267(2)$ & $-1.2273$ & 0.05(2) \\
                     &             & 12 & $A_1$ & $-1.1132(3)$ & $-1.1137$ & 0.04(3) \\
        \cmidrule{2-7}
                     & \multirow{3}{*}{(7, 7, 7/8)} & 4 & $A_1$ & $-1.3234(1)$ & $-1.3234$ & 0.00(1) \\
                     &             & 8 & $A_1$ & $-0.9776(4)$ & $-0.9739$ & -0.38(4) \\
                     &             & 12 & $A_1$ & $-0.8151(5)$ & $-0.8086$ & $-0.81(6)$ \\
        \bottomrule
    \end{tabular}
    \label{tab:more_data}
\end{table*}

\twocolumngrid
\subsection{Efficient imaginary-time evolution} \label{app:cpmc_fast_updates}
We perform all CPMC calculations with a development version of the \textsc{trot} Python package \cite{ad_afqmc_code}. We use a time step of $\Delta \tau = 0.005$ and 200-400 walkers, having verified for select cases that the energies agree within statistical error with $\Delta \tau = 0.001$ and 600 walkers. Measurements are taken after the propagation has equilibrated.

In \textsc{trot}, imaginary-time evolution and importance sampling are implemented in an algorithm that scales as $\mathcal{O}(N_s^3)$ per time step, where $N_s$ is the number of sites. Recall that the spin decomposition due to Hirsch \cite{hirsch_discrete_1983,hirsch_two-dimensional_1985} for the Hubbard interaction on site $i$ is given by \eqref{hs}
\begin{align}
    \hat{B}_i &= e^{-\Delta \tau U \hat{n}_{i \uparrow} \hat{n}_{i \downarrow}} \nonumber \\
    &= e^{-\Delta \tau U (\hat{n}_{i\uparrow} + \hat{n}_{i\downarrow}) / 2} \sum_{x_i = \pm 1} p(x_i) e^{\gamma x_i (\hat{n}_{i\uparrow} - \hat{n}_{i\downarrow})} \nonumber 
    \\
    &= \sum_{x_i = \pm  1} p(x_i) e^{\lambda_{\uparrow}(x_i) \hat{n}_{i \uparrow}} \cdot e^{\lambda_{\downarrow}(x_i) \hat{n}_{i \downarrow}},
\end{align}
where we defined $\lambda_{\uparrow / \downarrow}(x_i) = -\Delta \tau U/2 \pm \gamma x_i$ and $\gamma$ is determined by $\cosh \gamma = e^{\Delta \tau U/2}$. Using the fact that $\hat{n}_{i \sigma}^k = \hat{n}_{i \sigma}$, we can simplify
\begin{align}
    e^{\lambda_{\sigma}(x_i) \hat{n}_{i \sigma}} &= \sum_{k=0} \frac{1}{k!} [\lambda_{\sigma}(x_i)]^k \hat{n}_{i \sigma}^k \nonumber \\
    &= 1 + \hat{n}_{i \sigma} \left(e^{\lambda_{\sigma}(x_i)} - 1\right), 
\end{align}
which gives 
\begin{widetext}
    \begin{align}
        \hat{B}_i 
        &= \sum_{x_i = \pm  1} p(x_i) \left\{ 1 + \left(e^{\lambda_{\uparrow}(x_i)} - 1\right) \hat{n}_{i \uparrow} + \left(e^{\lambda_{\downarrow}(x_i)} - 1\right) \hat{n}_{i \downarrow} + \left(e^{\lambda_{\uparrow}(x_i)} - 1\right) \left(e^{\lambda_{\downarrow}(x_i)} - 1\right) \hat{n}_{i \uparrow} \hat{n}_{i \downarrow} \right\}. \label{propagator}
    \end{align}
\end{widetext}

To implement importance sampling and the positivity constraint, we sample the auxiliary fields $x_i$ from the weighted probability distribution 
\begin{align}
    \tilde{p}(x_i) \propto p(x_i) \max\left( 0, \frac{\braket{\psi_T | \hat{B}_i | \phi}}{\braket{\psi_T | \phi}} \right). 
\end{align}
Intuitively, the \emph{overlap ratio} $\braket{\psi_T | \hat{B}_i | \phi} / \braket{\psi_T | \phi}$ between a trial state $\ket{\psi_T}$ and walker $\ket{\phi}$ guides the random walk towards regions where the overlap with $\ket{\psi_T}$ is large. If $\ket{\psi_T}$ is a single Slater determinant, the overlap is given by
\begin{align}
    \braket{\psi_T | \phi} = \det\left[ \bm{\psi}_T^\dagger \bm{\phi} \right] = \det\left[ \mathbf{O}(\phi) \right],
\end{align}
where $\mathbf{O}(\phi) = \bm{\psi}_T^\dagger \bm{\phi}$ is the \emph{overlap matrix} and $\bm{\psi}_T, \bm{\phi}$ are coefficient matrices associated with $\ket{\psi_T}, \ket{\phi}$. For a $N_e$-electron system and GHF-type determinants, the matrices $\bm{\psi}_T, \bm{\phi}$ have shape $2N_s \times N_e$, implying that $\mathbf{O}(\phi)$ has shape $N_e \times N_e$. Thus, for a fixed filling factor, the evaluation of the overlap requires $\mathcal{O}(N_e^3) \sim \mathcal{O}(N_s^3)$ time to compute the determinant $\det\left[ \mathbf{O}(\phi) \right]$.

However, this can be sped up if one has access to the Green's function 
\begin{align}
    \frac{\braket{\psi_T | \hat{c}_{i \sigma}^\dagger \hat{c}_{j \tau} | \phi}}{\braket{\psi_T | \phi}} = G(\phi)_{ij}^{\sigma \tau} = \left[ \bm{\phi} \mathbf{O}(\phi)^{-1} \bm{\psi}_T^\dagger \right]_{ji}^{\tau\sigma}. \label{greens}
\end{align}
To see this, we insert \eqref{propagator} into the overlap ratio (assuming we've sampled a value for $x_i$) and obtain
\begin{align}
    &\frac{\braket{\psi_T | \hat{B}_i | \phi}}{\braket{\psi_T | \phi}} \nonumber \\
    &= 1 + \left(e^{\lambda_{\uparrow}(x_i)} - 1\right) \frac{\braket{\psi_T | \hat{n}_{i \uparrow} | \phi}}{\braket{\psi_T | \phi}} \nonumber \\
    &\qquad + \left(e^{\lambda_{\downarrow}(x_i)} - 1\right) \frac{\braket{\psi_T | \hat{n}_{i \downarrow} | \phi}}{\braket{\psi_T | \phi}}  \nonumber \\ 
    &\qquad + \left(e^{\lambda_{\uparrow}(x_i)} - 1\right) \left(e^{\lambda_{\downarrow}(x_i)} - 1\right) \frac{\braket{\psi_T | \hat{n}_{i \uparrow} \hat{n}_{i \downarrow} | \phi}}{\braket{\psi_T | \phi}}.
\end{align}
Applying the generalized Wick's theorem \cite{hendekovic_generalization_1981},
\begin{align}
    \frac{\braket{\psi_T | \hat{n}_{i \sigma} | \phi}}{\braket{\psi_T | \phi}} &= \frac{\braket{\psi_T | \hat{c}_{i \sigma}^\dagger \hat{c}_{i \sigma} | \phi}}{\braket{\psi_T | \phi}} = G(\phi)^{\sigma \sigma}_{ii} \\
    \frac{\braket{\psi_T | \hat{n}_{i \uparrow} \hat{n}_{i \downarrow} | \phi}}{\braket{\psi_T | \phi}} &= \frac{\braket{\psi_T | \hat{c}_{i \uparrow}^\dagger \hat{c}_{i \uparrow} \hat{c}_{i \downarrow}^\dagger \hat{c}_{i \downarrow} | \phi}}{\braket{\psi_T | \phi}} \nonumber \\
    &= G(\phi)^{\uparrow \uparrow}_{ii} G(\phi)^{\downarrow \downarrow}_{ii} - G(\phi)^{\uparrow \downarrow}_{ii} G(\phi)^{\downarrow \uparrow}_{ii},
\end{align}
and rearranging terms, we end up with
\begin{widetext}
    \begin{align}
        \frac{\braket{\psi_T | \hat{B}_i | \phi}}{\braket{\psi_T | \phi}}
        &= \left\{ 1 + \left(e^{\lambda_{\uparrow}(x_i)} - 1\right) G(\phi)^{\uparrow \uparrow}_{ii}\right\} \left\{ 1 + \left(e^{\lambda_{\downarrow}(x_i)} - 1\right) G(\phi)^{\downarrow \downarrow}_{ii} \right\} - \left(e^{\lambda_{\uparrow}(x_i)} - 1\right) \left(e^{\lambda_{\downarrow}(x_i)} - 1\right) G(\phi)^{\uparrow \downarrow}_{ii} G(\phi)^{\downarrow \uparrow}_{ii},
    \end{align}
\end{widetext}
which only requires $\mathcal{O}(1)$ time to retrieve the appropriate matrix elements of $\mathbf{G}(\phi)$. 

The evaluation of $\mathbf{G}(\phi)$ in \eqref{greens} involves matrix multiplications and an inversion, with cost scaling as $\mathcal{O}(N_e^3) \sim \mathcal{O}(N_s^3)$. If one were to recompute $\mathbf{G}(\phi)$ to use in importance sampling after each successive application of $\hat{B}_i$ over all sites, the total cost of the two-body propagation $\prod_{i=1}^{N_s} \hat{B}_i$ for \emph{one time step} would be $\mathcal{O}(N_s^4)$. We avoid this by updating the Green's function efficiently between consecutive applications of $\hat{B}_i$ with cost scaling as $\mathcal{O}(N_s^2)$. 

Consider the updated Green's function
\begin{align}
    G'(\phi)_{pq}^{\sigma \tau} &= \frac{\braket{\psi_T | \hat{c}_{p \sigma}^\dagger \hat{c}_{q \tau} \hat{B}_i | \phi}}{\braket{\psi_T | \hat{B}_i | \phi}} \nonumber \\
    &= \frac{\braket{\psi_T | \hat{c}_{p \sigma}^\dagger \hat{c}_{q \tau} \hat{B}_i | \phi}}{\braket{\psi_T | \phi}} \times \frac{\braket{\psi_T | \phi}}{\braket{\psi_T | \hat{B}_i | \phi}}.
\end{align}
The first term can be evaluated using Wick's theorem again, while the second term is the inverse of the overlap ratio. Repeating the exercise of inserting \eqref{propagator} into the first term (and dropping the dependence on $\phi$ from $\mathbf{G}$), we have
\begin{widetext}
    \begin{align}
        \frac{\braket{\psi_T | \hat{c}_{p \sigma}^\dagger \hat{c}_{q \tau} \hat{B}_i | \phi}}{\braket{\psi_T | \phi}}
        &= G_{pq}^{\sigma \tau} \frac{\braket{\psi_T | \hat{B}_i | \phi}}{\braket{\psi_T | \phi}} + \left(e^{\lambda_{\uparrow}(x_i)} - 1\right) G_{pi}^{\sigma \uparrow} \left\{ \left(e^{\lambda_{\downarrow}(x_i)} - 1\right) \left[ G_{ii}^{\uparrow \downarrow} \mathcal{G}_{iq}^{\downarrow \tau} - G_{ii}^{\downarrow \downarrow} \mathcal{G}_{iq}^{\uparrow \tau} \right] -\mathcal{G}_{iq}^{\uparrow \tau} \right\} \nonumber \\
        &\qquad \qquad \qquad \qquad + \left(e^{\lambda_{\downarrow}(x_i)} - 1\right) G_{pi}^{\sigma \downarrow} \left\{ \left(e^{\lambda_{\uparrow}(x_i)} - 1\right) \left[ G_{ii}^{\downarrow \uparrow} \mathcal{G}_{iq}^{\uparrow \tau} - G_{ii}^{\uparrow \uparrow} \mathcal{G}_{iq}^{\downarrow \tau} \right] - \mathcal{G}_{iq}^{\downarrow \tau} \right\},
    \end{align}
\end{widetext}
where we defined $\mathcal{G}_{pq}^{\sigma \tau} = G_{pq}^{\sigma \tau} - \delta_{pq} \delta_{\sigma \tau} = \left[ \mathbf{G} - \mathbf{I} \right]_{pq}^{\sigma \tau}$. 

Putting it all together, the updated Green's function can be calculated with 
\begin{widetext}
    \begin{align}
        G_{pq}^{'\sigma \tau} &= G_{pq}^{\sigma \tau} + \frac{e^{\lambda_{\uparrow}(x_i)} - 1}{O_\text{ratio}} \ G_{pi}^{\sigma \uparrow} \left\{ \left(e^{\lambda_{\downarrow}(x_i)} - 1\right) \left[ G_{ii}^{\uparrow \downarrow} \mathcal{G}_{iq}^{\downarrow \tau} - G_{ii}^{\downarrow \downarrow} \mathcal{G}_{iq}^{\uparrow \tau} \right] -\mathcal{G}_{iq}^{\uparrow \tau} \right\} \nonumber \\
        &\quad \qquad \qquad + \frac{e^{\lambda_{\downarrow}(x_i)} - 1}{O_\text{ratio}} \ G_{pi}^{\sigma \downarrow} \left\{ \left(e^{\lambda_{\uparrow}(x_i)} - 1\right) \left[ G_{ii}^{\downarrow \uparrow} \mathcal{G}_{iq}^{\uparrow \tau} - G_{ii}^{\uparrow \uparrow} \mathcal{G}_{iq}^{\downarrow \tau} \right] - \mathcal{G}_{iq}^{\downarrow \tau} \right\}, \label{update_green_2}
    \end{align}
\end{widetext}
where we introduced the shorthand $O_\text{ratio} = \braket{\psi_T | \hat{B}_i | \phi}/\braket{\psi_T | \phi}$. Note that $\mathbf{G}'$ is determined solely from the current Green's function $\mathbf{G}$ since $O_\text{ratio}$ also depends only on $\mathbf{G}$. To update each matrix element, the cost is $\mathcal{O}(1)$ since \eqref{update_green_2} can be evaluated by retrieving the appropriate matrix elements of $\mathbf{G}$. The cost per time step to update the entire Green's function is thus $\mathcal{O}(N_s^2)$ since $\mathbf{G}'$ is a $N_s \times N_s$ matrix.

\begin{figure}[b]
    \centering
    \includegraphics[width=0.9\linewidth]{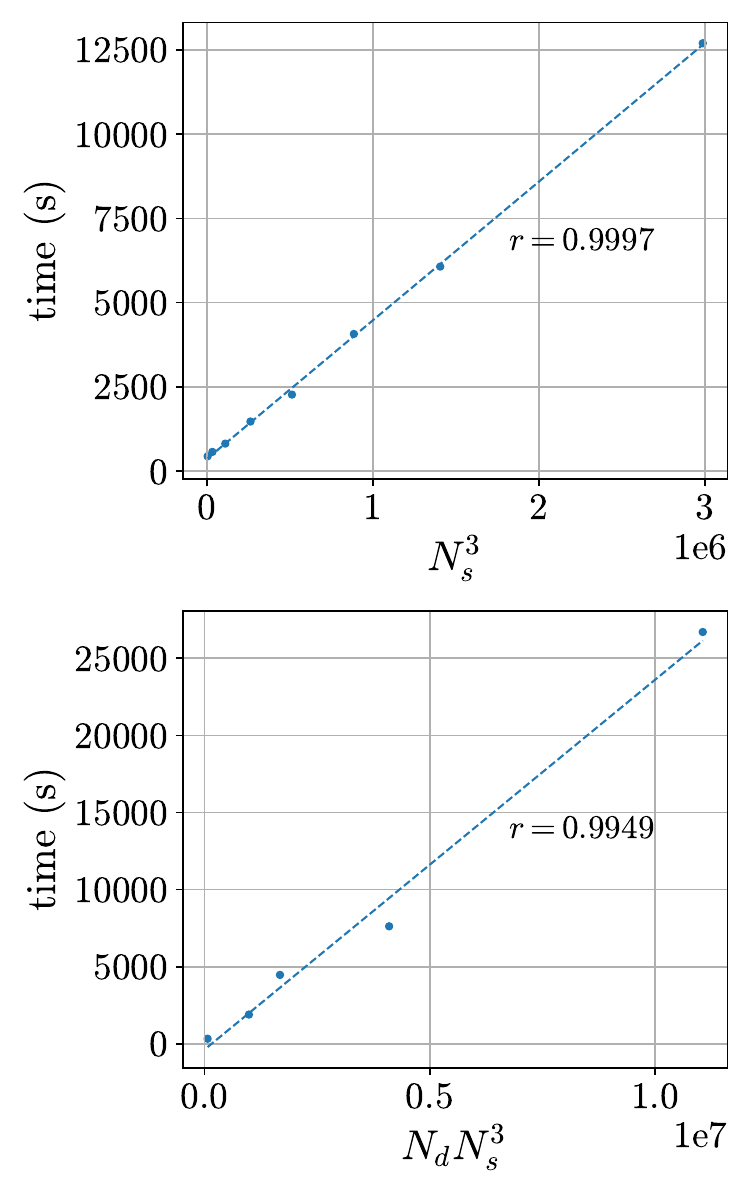}
    \caption{(Top) CPMC runtime scaling without symmetry projection. Calculations are performed on YC4 cylinders at quarter-filling ($\nu = 1/2$) and $U = 12$. (Bottom) CPMC runtime scaling with symmetry projection. Calculations are performed on XC4 cylinders at half-filling ($\nu = 1$) and $U = 8$, using only $S^2$ symmetry projection. The timings exhibit cubic scaling with the number of sites $N_s$, while symmetry projection introduces an additional linear scaling with the number of determinants $N_d$. The Pearson correlation coefficient $r$ of the linear fit is also shown.}
    \label{fig:timings}
\end{figure}

Therefore, the theoretical scaling of our CPMC algorithm is determined by the calculation of $\mathbf{G}$ between successive time steps, which scales as $\mathcal{O}(N_s^3)$. We show this empirically in Fig.~\ref{fig:timings}.

\subsection{Symmetry-projected trial wave functions} \label{app:symm_trial}
The projectors $\hat{\mathcal{P}}^s_{mm'}$, $\hat{\mathcal{P}}_\alpha$, and $\hat{\mathcal{P}}_K$ are defined in the main text Sec.~\ref{sec:symm_proj_trial}. In all cases considered in this work, we target the singlet spin sector ($s=0$, $m=m'=0$), for which the Wigner $D$-matrix reduces to $D^0_{00}(\Omega) = 1$ and the spin projector simplifies to
\begin{equation}\label{eq:singlet_proj}
  \hat{\mathcal{P}}^{s=0} = \frac{1}{8\pi^2}\int d\alpha\, d\beta\, d\gamma\, \sin\beta\ \hat{R}(\alpha,\beta,\gamma),
\end{equation}
where $\hat{R}(\alpha,\beta,\gamma) = e^{-i\alpha \hat{S}_z} e^{-i\beta \hat{S}_y} e^{-i\gamma \hat{S}_z}$. We discretize the $\alpha$, $\gamma$ integrations with the midpoint rule and the $\beta$ integration with Gauss-Legendre quadrature.

We construct the trial wave function by variationally optimizing a GHF determinant $|\psi_\text{GHF}\rangle$ in the presence of all three projectors, following the variation-after-projection (VAP) approach. The projected energy
\begin{equation}
  E = \frac{\langle\psi_\text{GHF}|\hat{H}\hat{\mathcal{P}}|\psi_\text{GHF}\rangle}
           {\langle\psi_\text{GHF}|\hat{\mathcal{P}}|\psi_\text{GHF}\rangle},
  \qquad
  \hat{\mathcal{P}} = \hat{\mathcal{P}}^{s=0}\hat{\mathcal{P}}_\alpha
                       \hat{\mathcal{P}}_K,
\end{equation}
is minimized with respect to the GHF orbital coefficients using L-BFGS, with gradients obtained by automatic differentiation. The optimization is repeated over multiple random initial perturbations of a converged GHF solution, and the trial with the lowest projected energy is selected.

The resulting trial is a linear combination of $N_d$ Slater determinants generated by the combined action of the three projectors on the optimized GHF reference. Within CPMC, each walker is a spin-collinear determinant with a definite $\hat{S}_z$ eigenvalue. Since the spin projector is Hermitian and commutes with the Hamiltonian, it can be applied to the walker rather than the trial when computing overlaps and local energies. The rightmost $\hat{S}_z$ rotation gives $e^{-i\gamma \hat{S}_z}|\phi\rangle = e^{-i\gamma m}|\phi\rangle$ with $m=0$, eliminating the $\gamma$ integration and reducing $N_d$ by a factor of $n_\gamma$. Each determinant in the expansion is then treated as a separate component of a multideterminant trial; the Green's function for each is updated using the fast update procedure described in App.~\ref{app:cpmc_fast_updates}, giving an overall cost of $\mathcal{O}(N_d N_s^3)$ per time step. This scaling is confirmed empirically in Fig.~\ref{fig:timings}. The Euler-angle grid used in the CPMC trial expansion is chosen by evaluating the mixed energy estimator with the initial CPMC walkers on a sequence of progressively finer grids, then selecting the coarsest grid for which the energy is converged to within a specified tolerance. Typical grids consist of $5-10$ points per angle.

\subsection{Infinite-cylinder extrapolations} \label{app:cpmc_extrap}
The energies in Fig.~\ref{fig:doped_energies} are fitted to the form $\varepsilon(N_x) = \varepsilon_\infty + m/N_x$. For the YC4 and YC6 cylinders, the uncertainty in $\varepsilon_\infty$ is estimated by combining a statistical contribution $\sigma_\text{stat}$ (standard error of the fitted intercept) with a systematic contribution $\sigma_\text{sys}$ that quantifies sensitivity to the fitting window. Specifically, we repeat the fit using only the largest $k$ cylinder lengths, with $k \in \{3, 4, 6\}$ for YC4 and $k \in \{3, 4, 5\}$ for YC6, and obtain intercepts $\{\varepsilon_\infty^{(k)}\}$. We take the systematic uncertainty to be the maximum deviation of the windowed-fit intercepts from their median value $\tilde{\varepsilon}_\infty$,
\begin{align}
    \sigma_\text{sys} = \max_{k} \left| \varepsilon_\infty^{(k)} - \tilde{\varepsilon}_\infty \right|,
\end{align}
and $\sigma_\text{stat}$ to be the standard error of the fit for $\tilde{\varepsilon}_\infty$. The total uncertainty is computed as $\sigma_\text{tot} = \sqrt{\sigma_\text{stat}^2 + \sigma_\text{sys}^2}$ and we report $\varepsilon_\infty \simeq \tilde{\varepsilon}_\infty \pm \sigma_\text{tot}$.

For the YC12 cylinder, we repeat the fit only for $k \in \{2, 3\}$ (\textit{i.e.} using the two and three longest cylinder lengths). We report $\varepsilon_\infty \equiv \bar{\varepsilon}_\infty \pm \sigma_\text{tot}$, where $\bar{\varepsilon}_\infty = (\varepsilon_\infty^{(2)} + \varepsilon_\infty^{(3)})/2$; $\sigma_\text{stat}$ is obtained by propagating the standard errors of $\varepsilon_\infty^{(k)}$ and $\sigma_\text{sys} = \max_{k} | \varepsilon_\infty^{(k)} - \bar{\varepsilon}_\infty |$.

\section{GHF, ED, and DMRG computational details}
\subsection{GHF and ED} \label{app:ghf_ed}
GHF and ED calculations are carried out using \textsc{pyscf} \cite{sun_libcint_2015,sun_p_2018,sun_recent_2020}. GHF solutions are optimized through the co-iterative augmented Hessian (CIAH) second-order solver \cite{sun_co-iterative_2017,sun_general_2017}, with orbitals perturbed and re-optimized upon convergence to unstable solutions. We initialize the self-consistent field (SCF) procedure with a variety of initial guesses, including the Néel-ordered state as well as real and complex random density matrices sampled from the standard Gaussian distribution $\mathcal{N}(0, 1)$. The GHF solutions reported in the main text are the lowest-energy states obtained across these initial guesses.

On the other hand, ED solutions are obtained from \textsc{pyscf}'s full configuration interaction (FCI) solver, with multiple roots computed to ensure convergence to the correct ground state.

\subsection{DMRG} \label{app:dmrg}
DMRG calculations for the energy are performed with the \textsc{itensor} software library \cite{itensor}. The matrix product state is initialized with a staggered antiferromagnetic (AFM) state dressed with random noise. The bond dimension $D_s$ for the first 6 sweeps ($1 \leq s \leq 6$) is set to $D_s \in \{10, 20, 100, 200, 300, 500\}$, before increasing in a uniformly exponential manner following $D(n) = d a^n$, where $n \in \mathbb{Z}^+$ starts from $n = 1$ and $a$ is a parameter that controls the rate of increase of $D$ up to the maximum $D_\text{max}$ \cite{chen_quantum_2022}. We choose $d = 512, a = 2^{1/5}$, and $D_\text{max} = 8192$. To extrapolate the energy $\varepsilon$ to the zero truncation error limit $\xi \to 0$, we repeat each value of $D(n)$ for two full sweeps (\textit{i.e.} sweeps $s$ and $s+1$) for $s > 15$, resulting in a total of 32 sweeps. The energy from each second sweep is then linearly fitted to the form $\varepsilon(\xi) = \varepsilon_0 + m\xi$. 

The DMRG energies reported in Fig.~\ref{fig:doped_energies} are the zero-truncation-error intercepts $\varepsilon_0$, with uncertainties estimated following the procedure used for the infinite-cylinder extrapolations in App.~\ref{app:cpmc_extrap}. The total uncertainty $\sigma_\text{tot}$ includes both a statistical contribution $\sigma_\text{stat}$ (the standard error of the fitted intercept) and a systematic contribution $\sigma_\text{sys}$ that quantifies sensitivity to the fitting window. To estimate $\sigma_\text{sys}$, we repeat the fit using only the smallest $k$ truncation error points with $k \in \{2, 3, 4\}$ or $k \in \{3, 4, 5\}$, obtaining intercepts $\{\varepsilon_0^{(k)}\}$. We define $\tilde{\varepsilon}_0 \equiv \text{median}_k \ \varepsilon_0^{(k)}$ and take
\begin{align}
    \sigma_\text{sys} = \max_{k} \left| \varepsilon_0^{(k)} - \tilde{\varepsilon}_0 \right|.
\end{align}
We take $\sigma_\text{stat}$ to be the standard error of the fit that gives $\tilde{\varepsilon}_0$ and report $\varepsilon_0 \equiv \tilde{\varepsilon}_0 \pm \sigma_\text{tot}$, with $\sigma_\text{tot} = \sqrt{\sigma_\text{stat}^2 + \sigma_\text{sys}^2}$.

On the other hand, DMRG calculations for the spin structure factor are performed using the \textsc{block2} code \cite{zhai_block2_2023} with full SU(2) spin symmetry.

\section{XC4 cylinder symmetries} \label{app:xc4_symm}
\subsection{$D_4$ symmetry group}
\begin{table}[H]
    \setlength{\belowcaptionskip}{10pt}
    \centering
    \caption{Symmetry operators of the XC4 lattice.}
    \begin{tabular}{c|l}
        \hline
        Symmetry operator & Action \\
        \hline
        $T_y$ & Translation by $L_y$ \\
        \hline
        \multirow{3}{*}{$G$} & Translation by $L_y/2$ followed by \\
        & a reflection along the vertical line \\
        & bisecting the lattice \\
        \hline
        \multirow{2}{*}{$R_{\pi}$} & Rotation by $\pi$ about the center \\
        & of the lattice \\
        \hline
    \end{tabular}
    \label{tab:xc4_symm}
\end{table}

The symmetry operators on XC4 cylinders are given in Table \ref{tab:xc4_symm} and satisfy the relations
\begin{gather}
    T_y = G^2 \\
    R_\pi G R_\pi = G^{-1} \\
    G^4 = T^2 = R_\pi^2 = 1 \label{xc4_cycle} \\
    [G, R_\pi] \neq 0,
    \label{commutation_xc_4xNx}
\end{gather}
which imply that the generators of the group are $G$ and $R_\pi$. The operators furnish a faithful 8-dimensional unitary \emph{permutation representation} (in the site basis) of the dihedral group $D_4$ (\textit{i.e.} $C_{4v}$), defined by
\begin{align}
    D_4 = \langle x, y : x^4 = y^2 = 1, xyx = y \rangle,
\end{align}
with group elements $\{1, x, x^2, x^3, y, yx, yx^2, yx^3\}$. An explicit mapping between group elements and symmetry operators may be obtained by identifying $x \equiv G, y \equiv R_\pi$. 

The conjugacy classes are shown in Table~\ref{tab:xc4_d4_class}, each of which can be identified with the canonical $D_4$ classification consisting of the identity ($E$), a $180^\circ$ rotation ($C_2$), $90^\circ$ rotations ($C_4$), and two distinct types of reflections ($\sigma_v, \sigma_d$). The character table, which we use to characterize the ground state symmetry, is given in Table~\ref{tab:xc4_d4_char_table}.
\begin{table}[H]
    \centering
    \caption{Conjugacy classes and character table of $D_4$.}
    \begin{minipage}[t]{0.48\linewidth}
        \centering
        \subcaption{Conjugacy classes.}
        \begin{tabular}{c|c}
            \hline
            Conjugacy & Canonical \\
            class & naming \\
            \hline
            $\{ 1 \}$ & $E$ \\
            $\{ G^2 \}$ & $C_2$ \\
            $\{ G, G^3 \}$ & $C_4$ \\
            $\{ R_\pi, R_\pi G^2 \}$ & $\sigma_v$ \\
            $\{ R_\pi G, R_\pi G^3 \}$ & $\sigma_d$ \\
            \hline
        \end{tabular}
        \label{tab:xc4_d4_class}
    \end{minipage} \hfill
    \begin{minipage}[t]{0.48\linewidth}
        \centering
        \subcaption{Character table.}
        \begin{tabular}{c|c|c|c|c|c}
            \hline
            Irrep & $E$ & $C_2$ & $C_4$ & $\sigma_v$ & $\sigma_d$ \\
            \hline
            $A_1$ & +1 & +1 & +1 & +1 & +1 \\
            $A_2$ & +1 & +1 & +1 & $-1$ & $-1$ \\
            $B_1$ & +1 & +1 & $-1$ & +1 & $-1$ \\
            $B_2$ & +1 & +1 & $-1$ & $-1$ & +1 \\
            $E$ & +2 & $-2$ & 0 & 0 & 0 \\
            \hline
        \end{tabular}
        \label{tab:xc4_d4_char_table}
    \end{minipage}
    \label{tab:xc4_d4_class_table}
\end{table}

\subsection{$D_2$ subgroup}
The $D_4$ group contains an Abelian $D_2$ subgroup with elements $\{1, G^2, R_\pi, R_\pi G^2\}$. Since the subgroup is Abelian, each element forms its own conjugacy class and has four one-dimensional irreps. These irreps provide a set of good quantum numbers for labeling eigenstates of $\hat{H}_1$ and defining a canonical basis inside degenerate subspaces.

\begin{table}[H]
    \centering
    \caption{Conjugacy classes and character table of $D_2$.}
    \begin{minipage}[t]{0.48\linewidth}
        \centering
        \subcaption{Conjugacy classes.}
        \begin{tabular}{c|c}
            \hline
            Conjugacy & Canonical \\
            class & naming \\
            \hline
            $\{ 1 \}$ & $E$ \\
            $\{ G^2 \}$ & $C_2$ \\
            $\{ R_\pi\}$ & $C_2^{'}$ \\
            $\{ R_\pi G^2 \}$ & $C_2^{''}$ \\
            \hline
        \end{tabular}
        \label{tab:xc4_d2_class}
    \end{minipage} \hfill
    \begin{minipage}[t]{0.48\linewidth}
        \centering
        \subcaption{Character table.}
        \begin{tabular}{c|c|c|c|c}
            \hline
            Irrep & $E$ & $C_2$ & $C_2^{'}$ & $C_2^{''}$ \\
            \hline
            $A_1$ & +1 & +1 & +1 & +1\\
            $A_2$ & +1 & +1 & $-1$ & $-1$ \\
            $B_1$ & +1 & $-1$ & +1 & $-1$ \\
            $B_2$ & +1 & $-1$ & $-1$ & +1 \\
            \hline
        \end{tabular}
        \label{tab:xc4_d2_char_table}
    \end{minipage}
    \label{tab:xc4_d2_class_table}
\end{table}

\nocite{REVTEX41Control}
\bibliographystyle{apsrev4-1}
\bibliography{ref_shufay, ref_code}
\end{document}